\documentclass[11pt]{article}
\usepackage{mathtools}
\usepackage{amssymb}
\usepackage{dsfont}
\usepackage{slashed}
\usepackage{fullpage}
\usepackage[colorlinks=true,linkcolor=blue,citecolor=magenta,linktocpage=true]{hyperref}

\usepackage{titlesec}
\titleformat*{\section}{\normalsize\bfseries}
\titleformat*{\subsection}{\normalsize\bfseries}
\titleformat*{\subsubsection}{\normalsize\bfseries}
\usepackage{cite}
\usepackage{comment}

\DeclareMathAlphabet{\bbvar}{U}{BOONDOX-ds}{m}{n}

\makeatletter
\renewcommand{\@dotsep}{10000}
\makeatother

\def\be#1\ee{\begin{align}#1\end{align}}

\def\rm#1{\mathrm{#1}}

\def\M{\mathcal{M}}

\def\rd{\textrm{d}}

\def\beq{\begin{eqnarray}}
\def\eeq{\end{eqnarray}}
\def\be{\begin{equation}}
\def\ee{\end{equation}}

\renewcommand{\theequation}{\thesection.\arabic{equation}}\numberwithin{equation}{section}

\renewpagestyle{plain}{
\sethead{}{}{YITP-21-72, IPMU21-0046}
\setfoot[\thepage][][]{}{\thepage}{}
}

\begin{document}
\thispagestyle{plain}

\title{\Large{\textbf{\sffamily Disformal map and Petrov classification \\ in modified gravity}}}

\author{\sffamily Jibril Ben Achour$^{1}$, \; Antonio De~Felice$^{2}$, \; Mohammad Ali Gorji$^{2}$, \\ \sffamily Shinji Mukohyama$^{2,3}$, \; Masroor C.~Pookkillath$^{2}$}
\date{\small{\textit{
$^{1}$Arnold Sommerfeld Center for Theoretical Physics, Munich, Germany \\ 
$^{2}$Center for Gravitational Physics, Yukawa Institute for Theoretical Physics, Kyoto University, 606-8502, Kyoto, Japan\\
$^3$Kavli Institute for the Physics and Mathematics of the Universe (WPI), The University of Tokyo, Kashiwa, Chiba 277-8583, Japan~
}}}

\maketitle

\hrule
\hspace{.5cm}

\begin{abstract}
Disformal transformation provides a map relating different scalar-tensor and vector-tensor theories and gives access to a powerful solution-generating method in modified gravity. In view of the vast family of new solutions one can achieve, it is crucial to design suitable tools to guide their construction. In this work, we address this question by revisiting the Petrov classification of disformally constructed solutions in modified gravity theories. We provide close formulas which relate the principal nulls directions as well as the Weyl scalars before and after the disformal transformation. These formulas allow one to capture if and how the Petrov type of a given seed geometry changes under a disformal transformation. Finally, we apply our general setup to three relevant disformally constructed solutions for which the seeds are respectively homogeneous and isotropic, static spherically symmetric and stationary axisymmetric. For the first two cases, we show that the Petrov type O and Petrov type D remain unchanged after a disformal transformation while we show that disformed Kerr black hole is no longer of type D but of general Petrov type I. The results presented in this work should serve as a new
toolkit when constructing and comparing new disformal solutions in modified gravity.
\end{abstract}
\hspace{.5cm}
\hrule

\newpage
\setcounter{page}{2}

\tableofcontents
\vspace{0.7cm}
\hrule

\newpage

\section{Introduction}\label{introduction}

The recent advent of gravitational wave astronomy~\cite{TheLIGOScientific:2016src,TheLIGOScientific:2017qsa,Monitor:2017mdv,LIGOScientific:2019fpa} as well as the progress in imaging the close environment of black holes~\cite{Akiyama:2019cqa,Akiyama:2019eap} have opened a new window to explore gravitational physics. These new observational channels on astrophysical ultra compact objects provide key tests of General Relativity (GR) in the strong field regime while constraining the different theories of modified gravity introduced so far~\cite{Yunes:2013dva,Berti:2015itd,Yunes:2016jcc, Perkins:2021mhb}. In particular, the on-going observational campaigns might allow us to test the fundamental no-hair theorem of GR and severely constrain the landscape of hairy black hole solutions found in modified gravity theories~\cite{Yunes:2016jcc, Cunha:2019ikd,Khodadi:2020jij}.

Among the different extensions of GR studied so far, the ones involving additional scalar and/or vector modes have been explored the most. The criterion of degeneracy of the Lagrangian introduced in~\cite{Langlois:2015cwa} and further studied in a series of works~\cite{Langlois:2015skt, Achour:2016rkg, BenAchour:2016fzp} has allowed a systematic construction of these higher order theories, revealing a new landscape of healthy theories dubbed Degenerate Higher Order Scalar-Tensor (DHOST) theories in the scalar-tensor framework and generalized degenerate Proca in the vector-tensor context~\cite{Heisenberg:2016eld, Jimenez:2016upj, Kimura:2016rzw, Heisenberg:2017mzp,Jimenez:2019hpl}. See~\cite{Kobayashi:2019hrl, Heisenberg:2018vsk, Langlois:2018dxi, Langlois:2020xbc} for reviews presenting a unifying point of view on the construction of these theories and their mixed scalar-vector-tensor versions. While the field equations of these degenerate theories are usually of order higher than two, they remain free from the Ostrogradsky ghost and provide the right number of physical degrees of freedom~\cite{Ganz:2020skf}. Moreover, the landscape of degenerate theories can be split into different subclasses which can be viewed as equivalence classes of theories under general invertible disformal field redefinitions. For non-invertible disformal transformations, new theories can be found~\cite{Deruelle:2014zza,Firouzjahi:2018xob}. In the case of scalar-tensor theories, mimetic theories~\cite{Chamseddine:2013kea} arise which are pathological in general~\cite{Firouzjahi:2017txv,Zheng:2017qfs,Hirano:2017zox,Gorji:2017cai,Gorji:2019rlm, BenAchour:2017ivq, Yoshida:2018kwy}. For the vector-tensor case, depending on the form of the disformal transformation, either Einstein-Aether model or gauge field mimetic scenario can be achieved which are free of disastrous pathologies~\cite{Gorji:2018okn,Jirousek:2018ago,Gorji:2019ttx}. Thus, disformal transformation provides the key tool to find new theories and to explore how physical properties of the gravitational field change when turning on different higher order scalar-tensor and vector-tensor interaction terms~\cite{Zumalacarregui:2013pma, Domenech:2015hka, Watanabe:2015uqa, Naruko:2020oha, Domenech:2019syf,Deffayet:2020ypa, Minamitsuji:2020jvf}.

In order to confront such theories of modified gravity with astrophysical observations, it is useful to provide exact solutions describing the gravitational field around rotating compact objects and to extract physical predictions. The search for such exact solutions has been intensively developed over the last decades and is currently actively pursued. In the scalar-tensor framework, it was understood already some time ago that the powerful no-hair theorem of GR can be extended in different ways, providing a key guidance in exploring the solution space of these theories (see~\cite{Volkov:2016ehx, Herdeiro:2015waa, Lehebel:2018zga} for reviews). As a consequence, a whilst of new exact solutions have been explored during the last years. A starting point for this exploration has been the construction of stealth solutions~\cite{ArkaniHamed:2003uy,Mukohyama:2005rw,Babichev:2013cya}, for which the additional modes do not gravitate at the background level so that their whole effects on the background is to change the cosmological constant. In this respect, the stealth solutions mimic exactly GR background solutions while revealing the presence of the additional degrees of freedom at the level of perturbations. General conditions for the existence of such stealth solutions have been classified for DHOST theories, delimiting thus an interesting region of the solution space~\cite{Kobayashi:2014eva, Babichev:2017guv, BenAchour:2018dap, Motohashi:2018wdq, Minamitsuji:2018vuw, Bernardo:2019yxp, Motohashi:2019sen, Takahashi:2020hso, Charmousis:2019vnf}. See~\cite{Cisterna:2016nwq, Chagoya:2016aar, Chagoya:2017ojn, Heisenberg:2017xda, Babichev:2017rti, Heisenberg:2017hwb, Kase:2018voo, Kase:2018owh, Minamitsuji:2021gcq} for exact black holes solutions in vector-tensor theories. The exploration of this solution space beyond the stealth sector has then been addressed by several means. Among them, the disformal solution-generating method discussed in~\cite{BenAchour:2019fdf} provides a simple and powerful tool to obtain non-stealth exact solutions. Given a seed solution with a non-trivial scalar and/or vector profiles, one can straightforwardly build up a new exact solution of another scalar-tensor/vector-tensor theory. If the seed theory is degenerate, so will be the target theory as far  as the disformal map is invertible. This method was recently used to construct the first non-stealth disformed Kerr black hole solution for DHOST theories~\cite{BenAchour:2020fgy, Anson:2020trg}, providing thus an interesting arena to confront these theories with current observations. See also~\cite{Baake:2021kyg} for a more recent application of this solution-generating method to higher dimensional rotating black holes.

With this vast landscape of exact solutions, it becomes crucial to find methods to classify them and characterize their physical properties. To a large extent, most of the investigations have focused on perturbative analysis of black hole and cosmological stealth solutions, revealing crucial information on the quasinormal modes as well as the issues of the strong coupling~\cite{Ogawa:2015pea, deRham:2019gha, Takahashi:2019oxz, Motohashi:2019ymr,Khoury:2020aya,Gorji:2020bfl, Charmousis:2019fre, Bernardo:2020ehy, Tomikawa:2021pca, Langlois:2021aji, Takahashi:2021bml}. However, some key information about these solutions can be revealed only by mean of non-perturbative methods, such as the nature of their principal null directions (PNDs), which in turn, is intimately related to the non-perturbative gravitational waves carried in the non-stationary regime. Such information can be obtained through the well known Petrov classification. In view of the disformal method used to construct these new solutions, an interesting question is whether the Petrov type of a seed solution changes under a disformal transformation or not. It is also intriguing to ask how the properties of the PNDs of the seed change when going from the Einstein frame to the Jordan frame. The goal of this work is to provide the required tools to answer these questions and allows one to study non-perturbative features of the known exact solutions in modified gravity theories.

Introduced in the mid fifties, the Petrov classification characterizes kinematical properties of the gravitational fields using the algebraic properties of the Weyl tensor. In the so-called Newman-Penrose formalism, the information encoded in the ten independent components of the Weyl tensor are translated in terms of the five complex scalars known as Weyl scalars which represent the projection of the Weyl tensor in a special complex null basis built from the PNDs~\cite{Chandrasekhar:1985kt,Stephani:2003tm}. From the behaviour of the Weyl scalars under general Lorentz transformations, one can identify Lorentz-invariant combinations of them which then allow one to build a coordinate-independent and Lorentz-invariant algebraic classification of the underlying gravitational field. This method is by construction non-perturbative and purely kinematical, providing thus a powerful classification of gravitational fields. Since its introduction, the Petrov classification has been used as a guide in the search of exact solutions in GR. Indeed, by restricting from the beginning the solution to be algebraically special of a given Petrov type, one can dramatically simplify the problem. From that perspective, providing a clear understanding on how the Petrov type of a given seed is modified through a disformal solution-generating mapping and providing close formulas to identify the algebraic type of the target solution allow us to generalize this approach to modified gravity. Moreover, the knowledge of the algebraic type is in close relation with the gravitational waves content of a given solution, and powerful theorems encoding their propagation properties that have been derived in GR such as the peeling theorem. By investigating the Petrov type of disformal solution, one could therefore find some interesting ways to generalize these GR theorems to modified gravity. With these motivations in mind, we investigate in this paper a general method to systematically determine Petrov type of a spacetime before and after disformal transformation. This approach provides an efficient way to understand kinematical properties of the black hole solutions in modified gravity in a non-perturbative manner and without specifying any dynamics. From the solution-generating point of view, the results presented in the following shall serve both as a guide when constructing a given disformal exact solution and as a toolkit to compare different  exact disformally constructed solutions in modified gravity.

The paper is organized as follows. In Section~\ref{sec2}, we introduce a new map encoding the disformal transformation of the null tetrads in the Einstein and Jordan frames in the framework of the standard $2+2$ null decomposition. Using this map, we derive general relations relating the null directions in the Einstein and Jordan frames. In Section~\ref{sec3}, we provide a suitable decomposition of the disformed Weyl tensor under disformal map and discuss the symmetry properties of the different contributions. Using this result, we derive the relations between Weyl scalars in the Einstein and Jordan frames. This provides the main result of this work which allows to determine how the Petrov type of a seed geometry changes through a disformal mapping. In subsection~\ref{secgencond}, we provide close formula for the relations between the Weyl scalars in the case where the disformal parameter is small. Finally, in Section~\ref{sec4}, we apply our formalism to concrete examples. We show that the Petrov type O and Petrov type D remain unchanged for disformed homogeneous and isotropic and static spherically symmetric black hole solutions respectively while it changes from type D to type I for a disformed Kerr black hole. We comment on the mechanism at hand behind this result which is intimately related to the properties of scalar/vector profiles used in the construction of the new solution. Section~\ref{sec5} is devoted to the summary and discussion. Explicit formulas for the decomposition of the disformed Weyl scalars are presented in Appendix~\ref{app-tetrad-rep}.

\section{Disformal transformation of null directions}\label{sec2}

In this section, we briefly review the standard $2+2$ null decomposition of the spacetime geometry in term of null tetrads which provides the first step to develop the Newman-Penrose formalism. Then, we introduce the disformally transformed null tetrads which play a key role in our analysis. Finally, we provide explicit relations to derive the null directions in the Jordan frame from the ones in the Einstein frame and vice versa.

\subsection{$2+2$ null decomposition}

Consider a 4-dimensional Lorentzian manifold $({\cal M},{\tilde g})$. At any point of the spacetime, we can define an observer frame $\tilde{\theta}^a{}_{\mu}$, or tetrad in the following, which allows one to project the metric into the local rest frame of the observer as follows
\begin{equation}\label{g-tilde-T}
{\tilde g}_{\mu\nu} = \eta_{ab}\,{\tilde \theta}^a{}_{\mu} {\tilde \theta}^b{}_{\nu} \,, \hspace{1cm}
\eta_{ab} = {\tilde g}_{\mu\nu}{\tilde \theta}^{\mu}{}_a \, {\tilde \theta}^{\nu}{}_b \,,
\end{equation}
where $a,b=0,1,2,3$ are the Lorentz indices and $\eta_{ab}$ is the Minkowski metric associated to the local rest frame of the observer. The Newman-Penrose decomposition of the spacetime geometry requires a null tetrad basis. This tetrad (and tetrad inverse) components are given in term of four complex null vectors such that
\begin{equation}\label{T-tilde}
{\tilde \theta}^a{}_{\mu} = \big( - {\tilde k}_\mu, - {\tilde l}_\mu, \tilde{\bar m}_\mu, \tilde{ m}_\mu \big) \,, \hspace{1cm}
{\tilde \theta}^{\mu}{}_a = \big( {\tilde l}^\mu, {\tilde k}^\mu, \tilde{m}^\mu, \tilde{\bar m}^\mu \big) \,,
\end{equation}
where  ${\tilde m}$ and ${\tilde {\bar m}}$ are complex conjugate to each other, and the Minkowski metric takes the null form
\begin{eqnarray}\label{eta}
\eta_{ab} = \eta^{ab} = \begin{pmatrix}
0 & -1 & 0 & 0 \\
-1 & 0 & 0 & 0 \\
0 & 0 & 0 & 1 \\
0 & 0 & 1 & 0 
\end{pmatrix} \,.
\end{eqnarray}
The tetrad and  inverse tetrad components satisfy the orthogonality relations
\begin{equation}\label{T-complete}
{\tilde \theta}^a{}_{\mu} {\tilde \theta}^{\mu}{}_b = \delta^a{}_b \,, \hspace{1cm}
{\tilde \theta}^{\mu}{}_a {\tilde \theta}^a{}_{\nu} = \delta^\mu{}_\nu \,.
\end{equation}
Then, the metric components can be decomposed in terms of the tetrad components as
\begin{equation}\label{g-tilde}
{\tilde g}_{\mu\nu} = - {\tilde l}_\mu {\tilde k}_\nu - {\tilde k}_\mu {\tilde l}_\nu
+ {\tilde m}_\mu {\tilde{\bar m}}_\nu + {\tilde {\bar m}}_\mu {\tilde m}_\nu \,,
\end{equation}
where the null vectors satisfy the following normalization and orthogonality conditions
\begin{eqnarray}\label{null-tilde}
{\tilde g}_{\mu\nu}{\tilde l}^\mu{\tilde k}^\nu = -1 \,, \hspace{1cm}
{\tilde g}_{\mu\nu}{\tilde m}^\mu{\tilde{\bar m}}^\nu = 1 \,.
\end{eqnarray}
All other contractions between these four null vectors vanish.

The tangent space of the manifold ${\cal M}$ at point $p$ is spanned by the above tetrad components as $T_p({\cal M}) = \mbox{Span}({\tilde l},{\tilde k},{\tilde m},\tilde{\bar m})$. The 2-dimensional spacelike surface ${\cal S}$ with tangent space $T_p({\cal S}) = \mbox{Span}({\tilde m},\tilde{{\bar m}})$ constitutes null foliation of ${\cal M}$. The induced metric on this surface can be expressed in terms of its basis as ${\tilde q}_{\mu\nu} = {\tilde m}_\mu {\tilde{\bar m}}_\nu + {\tilde{\bar m}}_\mu {\tilde m}_\nu$. The projection from $T_p({\cal M})$ to $T_p({\cal S})$ then is given by
${\tilde q}^\mu{}_\nu = \delta^\mu{}_\nu + {\tilde l}^\mu {\tilde k}_\nu + {\tilde k}^\mu {\tilde l}_\nu $.  Note that, by definition 
\begin{eqnarray}\label{orthogomal-kl}
W_\mu {\tilde l}^\mu = 0 = W_\mu \tilde{k}^\mu \,,
\hspace{1.5cm} \forall\hspace{.2cm} W \in T_p^\star({\cal S}) \,,
\end{eqnarray}
where $W=W_\mu{d}x^\mu$ in coordinate basis of the cotangent space $T_p^\star({\cal S})$. On the other hand, the 2-dimensional quotient space ${\cal M}/{\cal S}$ is spanned by the null vectors ${\tilde l}$ and ${\tilde k}$ such that $T_p({\cal M}/{\cal S}) = \mbox{Span}({\tilde l},{\tilde k})$. On this 2-dimensional quotient space, one has
\begin{eqnarray}\label{orthogomal-mmbar}
W_\mu {\tilde m}^\mu = 0 = W_\mu \tilde{\bar m}^\mu \,,
\hspace{1.5cm} \forall \hspace{.2cm} W \in T_p^\star({\cal M}/{\cal S}) \,.
\end{eqnarray}

Having reviewed the standard $2+2$ decomposition of the spacetime, we now turn to the disformal transformation of the metric in terms of the null tetrad. These transformations will provide the key map between the null directions in the Jordan and Einstein frames. 

\subsection{Disformal transformation of null tetrad/vectors}

\label{map}

We now consider a modified theory of gravity on ${\cal M}$ which includes scalar and vector degrees of freedom in addition to the usual tensor degrees of freedom of GR encoded in the metric. In particular we consider two scalar fields $A$ and $B$ which live in ${\cal M}$ as well as a 1-form $V \in T^\star_p(\cal M)$. 

To be concrete, we denote ${\tilde g}$ the metric in the Jordan frame to which the matter is minimally coupled. From now on, any quantity with tilde is considered in the Jordan frame. Then, we assume that this metric can be obtained by a disformal transformation of the metric $g$ in the Einstein frame such that 
\begin{equation}\label{DT}
{\tilde g}_{\mu\nu} = A\, {g}_{\mu\nu} + B\, V_\mu V_\nu \,,
\end{equation}
where $V_\mu$ is the component of the 1-form $V$ so that $V = V_{\mu} \rd x^{\mu}$. In the particular case when we remove the transverse degrees of freedom such that $V_\mu = \partial_\mu \phi$, where $\phi$ is a scalar field, $A$ and $B$ are functions of $\phi$ and its kinetic energy $g^{\mu\nu}\partial_\mu\phi \partial_\nu\phi$, one recovers the well-known disformal transformation which is widely used in the context of the scalar-tensor theories~\cite{Zumalacarregui:2013pma}. Here, however, we keep the setup as general as possible and we do not impose this restriction. The inverse metric defined by ${\tilde g}^{\mu\nu} {\tilde g}_{\mu\rho}= \delta^\nu{}_\rho$ is given by
\begin{equation}\label{inv-DT-metric}
{\tilde g}^{\mu\nu} = \frac{1}{A} \Big[ {g}^{\mu\nu} - \frac{B}{A+B Y} 
(g^{\mu\alpha} V_\alpha) (g^{\nu\beta}V_\beta) \Big] \,, \hspace{1cm}
\text{with} \qquad Y = g^{\mu\nu}V_\mu V_\nu \,,
\end{equation}
where we have used that ${g}^{\mu\nu} {g}_{\mu\rho}= \delta^\nu{}_\rho$. As such, the disformal map~\eqref{DT} allows one to express the physics either in the Jordan or Einstein frames depending on the target. 

While the map~\eqref{DT} is a mere field redefinition in the absence of any matter field, it nevertheless allows one to explore the physics of modified theories of gravity in an efficient way. In particular, this field redefinition can be used as a powerful solution-generating method in modified gravity, generalizing the standard method based on conformal transformation.  For a theory defined by its action $S$ and field content $(g, A, B, V)$, knowing an exact solution of this theory allows one to construct straightforwardly a new exact solution for the theory defined by the action $\tilde{S}$ with field content $(\tilde{g}, A, B, V)$ where ${\tilde g}$ and ${g}$ are related to each other through the disformal map~\eqref{DT}. Then, assuming that matter fields minimally couple to the metric ${\tilde g}$ in the Jordan frame, they will non-minimally couple to the Einstein frame metric $g$, such that at the end, the physics turns out to be inequivalent to the case with matter minimally coupled to the Einstein frame metric. In particular, the observables associated to test probes propagating on these backgrounds will be different in general. This solution-generating method has been explored in~\cite{BenAchour:2019fdf} and further applied in recent works to construct exact rotating black hole solutions in DHOST theories~\cite{BenAchour:2020fgy, Anson:2020trg, Baake:2021kyg}. While some properties of these new solutions have already been discussed, a systematic approach would be desirable. In particular, the application of the Newman-Penrose formalism and the associated Petrov classification would allow one to characterize these new solutions in a  non-perturbative and systematic manner. The key tool of this approach being the use of a suitable null tetrad,  it is therefore mandatory, as a first step, to understand how such null frame changes under a disformal transformation.

Indeed, after performing the disformal transformation~\eqref{DT}, the disformal term $B$ changes the null directions in general while the conformal term preserves them. Consequently, the null directions $( \tilde{l}^\mu, \tilde{k}^\mu, {\tilde m}^\mu, \tilde{\bar m}^\mu )$ associated to the Jordan frame, that we presented in the previous subsection, are no longer null in the Einstein frame. Therefore, the first task is to find relation between the null directions $( \tilde{l}^\mu, \tilde{k}^\mu, {\tilde m}^\mu, \tilde{\bar m}^\mu )$ in the Jordan frame and the null directions $( l^\mu, k^\mu, m^\mu, {\bar m}^\mu ) $ in the Einstein frame. In the following, we introduce the key object to compute in a systematic manner these relations. Consider therefore the metric $g$ in the Einstein frame and let us expand it as
\begin{equation}\label{g-T}
{ g}_{\mu\nu} = \eta_{ab}\,{ \theta}^a{}_{\mu} { \theta}^b{}_{\nu} \,, \hspace{1cm}
\eta_{ab} = { g}_{\mu\nu}{ \theta}^{\mu}{}_a \, { \theta}^{\nu}{}_b \,,
\end{equation}
where
\begin{equation}\label{T}
{ \theta}^a{}_{\mu} = \big( - { k}_\mu, - { l}_\mu, {\bar m}_\mu, { m}_\mu \big) \,, \hspace{1cm}
{ \theta}^{\mu}{}_a = \big( { l}^\mu, { k}^\mu, {m}^\mu, {\bar m}^\mu \big) \,,
\end{equation}
denote the new tetrad basis and $(l_{\mu}, k_{\mu}, m_{\mu}, \bar{m}_{\mu})$ are the null directions in the Einstein frame. The metric in the Einstein frame splits as
\begin{equation}\label{g}
{ g}_{\mu\nu} = - { l}_\mu { k}_\nu - { k}_\mu { l}_\nu
+ { m}_\mu {\bar { m}}_\nu + {\bar { m}}_\mu { m}_\nu \,,
\end{equation}
where again we have normalized the null directions as
\begin{eqnarray}\label{null}
{ g}_{\mu\nu}{ l}^\mu{ k}^\nu = -1 \,, \hspace{1cm}
{ g}_{\mu\nu}{ m}^\mu{\bar{ m}}^\nu = 1 \,,
\end{eqnarray}
and have supposed that all other contractions between the four null vectors in the Einstein frame vanish.

The decomposition~\eqref{g} of the metric in terms of the null vectors provides an alternative way to write down the disformal transformation~\eqref{DT}. While it is standard to view it as a field redefinition between the metrics ${\tilde g}$ and $g$, one can equivalently understand it as a redefinition of the corresponding tetrad basis associated to the metric ${\tilde g}$ and $g$. This suggests to introduce the map $J^a{}_b(A, B, V)$ which relates the two tetrad basis in Einstein and Jordan frames as follows
\begin{equation}\label{T-map}
{\tilde \theta}^a{}_{\mu} \equiv J^a{}_b(A,B,V)\, {\theta}^b{}_{\mu} \,,
\end{equation}
which depends on the additional scalar and vector fields. Looking at the relation between the metric components and the tetrads in Eqs.~\eqref{g-tilde-T} and~\eqref{g-T}, the map can be thought of as the square root of the disformal transformation~\eqref{DT}. Physically, this map can be also interpreted as a change of local rest frame for an observer: namely ${\theta}^b{}_{\mu}$ corresponds to the local freely falling rest frame of the observer with respect to the Einstein frame while  ${\tilde \theta}^a{}_{\mu}$ represents its counterpart in the Jordan frame. Therefore, the matrix $J^a{}_b(A,B,V)$  encodes the local transformation one has to implement to relate the local freely falling observers in those frames. Equivalently, one can interpret the map $J$ as relating the two inequivalent realizations of the equivalence principle in the Einstein and Jordan frames. Notice that this map \textit{is not} a Lorentz transformation but it is defined up to Lorentz transformations which act on $J^a{}_b$ as $J^a{}_b = \Lambda^a{}_{a'} \Lambda^{b'}{}_b J^{a'}{}_{b'}$.

Moreover, besides the interpretation of the map J, it is important to stress that because each frames ${\theta}^b{}_{\mu}$ and ${\tilde \theta}^a{}_{\mu}$ can be associated to an observer, they implicitly introduce a notion of test matter field minimally coupled respectively to $g_{\mu\nu}$ and $\tilde{g}_{\mu\nu}$. When analyzing the properties of the two disformally related metrics, such as the decomposition of the Weyl tensor with respect to each frames, the presence of this disguised test matter field makes the two solutions inequivalent since one has to implicitly specify the coupling of this test matter field to each metric. This fact will be important when performing the Petrov classification of these two disformally related solutions.

Let us now find the explicit expression for the map $J^a{}_b$ in terms of the field content $(A, B, V)$. Writing the disformal transformation~\eqref{DT} in terms of the tetrad basis, we find
\begin{eqnarray}\label{Va}
\eta_{ab} {\tilde \theta}^a{}_\mu {\tilde \theta}^b{}_\nu = 
\Big[ A \eta_{ab} + B V_a V_b \Big] { \theta}^a{}_\mu { \theta}^b{}_\nu \,, \hspace{1cm}
V_a \equiv V_\alpha \theta^\alpha{}_a \,,
\end{eqnarray}
where $V_a $ are components of the vector field along the directions of the Einstein frame tetrad basis so that $V_\mu = V_a \theta^a{}_\mu$. Substituting the definition~\eqref{T-map} of $J^a{}_b$ in the above relation, one finds that
\begin{equation}\label{J}
J^a{}_b = \sqrt{A} \Big( \delta^a{}_b + \frac{\beta}{1-\beta Y} V^a V_b \Big) 
= \sqrt{A} \big( \delta_b{}^a + {\tilde \beta} {\tilde V}_b {\tilde V}^a \big) \,,
\end{equation}
up to a Lorentz transformation, where ${\tilde V}_a = V_\alpha {\tilde \theta}^\alpha{}_a$ is the tetrad component of $V_\mu$ in the Jordan frame tetrad basis and we have defined
\begin{equation}\label{beta}
{\tilde \beta} \equiv\frac{1}{\tilde Y} \bigg[\frac{1}{\sqrt{1-B{\tilde Y}}}-1\bigg] \,; \hspace{1cm}
\beta \equiv\frac{1}{Y} \bigg[1-\frac{\sqrt{A}}{\sqrt{A+BY}}\bigg] \,.
\end{equation}
Note that the denominator in~\eqref{J} is not singular since $\beta = Y^{-1}$ requires $A=0$ which is not allowed by the assumed invertibility of the disformal transformation (see \eqref{J-det} below). In the above equation we have also defined 
\begin{equation}\label{Y-tilde}
{\tilde Y} = {\tilde V}_a {\tilde V}^a = \frac{Y}{A+BY} \,; \hspace{1cm}
{\tilde V}_a \equiv \frac{V_a}{\sqrt{A+BY}} \,.
\end{equation}
We can proceed along the same line with the inverse transformation from the Einstein frame to the Jordan frame. We define the inverse map $T^a{}_b$ as follows
\begin{equation}\label{T-map-i}
{\theta}^a{}_\mu = T^a{}_b \, {\tilde \theta}^b{}_{\mu} \,; \hspace{1cm}
T^a{}_b \equiv \big(J^{-1}\big)^a{}_b \,.
\end{equation}
The inverse transformation matrix can be easily obtained by substituting Eq.~\eqref{J} into the relation $(J^{-1})^a{}_c J^c{}_b = \delta^a{}_b$ which gives
\begin{eqnarray}\label{J-inv}
T^a{}_b = \frac{1}{\sqrt{A}} \big( \delta^a{}_b - \beta V^a V_b \big)
= \frac{1}{\sqrt{A}} \Big( \delta^a{}_b 
- \frac{{\tilde \beta}}{1+{\tilde \beta}{\tilde Y}} {\tilde V}^a {\tilde V}_b \Big) \,.
\end{eqnarray}
Note that the invertibility of the map Eq.~\eqref{T-map} is guaranteed by demanding that the determinant of $J^a{}_b$ does not vanish. It provides the following condition
\begin{equation}\label{J-det}
J = T^{-1} = \frac{\sqrt{-{\tilde g}}}{\sqrt{-g}}= A^{3/2} (A+BY)^{1/2}  \neq 0 \,,
\end{equation}
where $J\equiv \mbox{det} (J^a{}_b)$ and $T \equiv \mbox{det} (T^a{}_b)$.

We have therefore achieved the first step: we have obtained the map $J^a{}_b$ (together with its inverse $T^a{}_b$), which allows one to write down explicitly the local projection of the tetrad basis ${\theta}^{\mu}{}_b$ in the Einstein frame onto the one of the Jordan frame, i.e.\ ${\tilde \theta}^{\mu}{}_a$, which can be summarized by the relations
\begin{equation}\label{T-map-E}
{\tilde \theta}^{\mu}{}_a \equiv T^b{}_a\, {\theta}^{\mu}{}_b \,, \hspace{1cm}
{\theta}^{\mu}{}_a \equiv J^b{}_a \, {\tilde \theta}^{\mu}{}_b \,.
\end{equation}
This map provides a key tool to write down explicit relations between the null directions in the Einstein frame and the ones in the Jordan frame. Using the explicit expressions for the map $J^a{}_b$ and its inverse, i.e.\ Eqs.~\eqref{J} and~\eqref{J-inv}, one obtains the relations
\begin{eqnarray}\label{T-dis}
	{\tilde \theta}^\mu{}_a = \frac{1}{\sqrt{A}} \,
	\Big( {\theta}^\mu{}_a - \beta\, V^\mu V_a \Big) \,, \hspace{1cm}
	{\theta}^\mu{}_a = \sqrt{A} \, \Big( {\tilde \theta}^\mu{}_a 
	+ {\tilde \beta}\, {\tilde V}^{\mu} {\tilde V}_a \Big) \,.
\end{eqnarray}
Substituting the definition of the tetrad in terms of the null directions, i.e.\ Eqs.~\eqref{T-tilde} and~\eqref{T}, in the first equation in~\eqref{T-dis}, we find the following mapping between the Jordan and Einstein null directions up to a Lorentz transformation
\begin{eqnarray}\label{kl-disform}
	{\tilde l}^\mu &=& \frac{1}{\sqrt{A}} \,
	\Big[ {l}^\mu{} - \beta (V_{\alpha}l^\alpha{}) (g^{\mu\beta} V_\beta) \Big] \,, \hspace{1.5cm}
	{\tilde k}^\mu = \frac{1}{\sqrt{A}} \,
	\Big[ {k}^\mu{} - \beta (V_{\alpha}k^\alpha{}) (g^{\mu\beta} V_\beta) \Big] \,,
	\nonumber \\
	{\tilde m}^\mu &=& \frac{1}{\sqrt{A}} \,
	\Big[ {m}^\mu{} - \beta (V_{\alpha}m^\alpha{}) (g^{\mu\beta} V_\beta) \Big] \,, \hspace{.8cm}
	{\tilde {\bar m}}^\mu = \frac{1}{\sqrt{A}} \,
	\Big[ {{\bar m}}^\mu{} - \beta (V_{\alpha}{\bar m}^\alpha{}) (g^{\mu\beta} V_\beta) \Big] .
\end{eqnarray}
The first two equations have already been obtained by the direct calculations for the case of the disformed Kerr black hole solution in the context of DHOST theories with $V_\mu=\partial_\mu\phi$~\cite{BenAchour:2020fgy}. In contrast, we provide here a systematic derivation of these key relations based on the introduction of the map $J^a{}_b$ for a more general setup. These relations provide a crucial tool to understand how the properties of null rays get modified under a disformal transformation, and how the causal structure of a given background changes through such mapping. 

Before closing this section, let us point out some useful consequences of these relations. Notice that after performing a disformal transformation, the null directions ${\tilde l}$ and ${\tilde k}$ do not change if $V \in T_p^\star({\cal S})$ as shown in~\eqref{orthogomal-kl} while the null directions ${\tilde m}$ and $\tilde{\bar m}$ do not change if $V \in T_p^\star({\cal M}/{\cal S})$ which can be seen from~\eqref{orthogomal-mmbar}. However, the null directions change in general for $V \in T_p^\star({\cal M})$. This observation will reveal useful when interpreting the results of the next sections.

We are now ready to investigate how the Petrov types change under a disformal map.

\section{Petrov classification}

\label{sec3}

As explained in the introduction, the Petrov classification is based on the algebraic properties of the Weyl tensor and allows one to characterize kinematical properties of the different gravitational fields in a coordinate-independent manner. There are several algorithms to determine the Petrov type of a given spacetime. We refer the readers to~\cite{Chandrasekhar:1985kt,Stephani:2003tm} for standard textbook reviews of the method as well as to~\cite{Coley:2007tp} for a more general perspective. In the following, we shall follow the approach based on the PNDs. Having determined in the previous section how the null directions transform under a disformal map, this approach will allow us to provide formula encoding how the Petrov type of a given disformally constructed solution is related to the Petrov type of the corresponding seed solution. 

\subsection{Decomposition of the Weyl tensor}

Consider a seed solution $(\M, g)$ in the Einstein frame and its disformal transformed solution $(\M, \tilde{g})$ in the Jordan frame.  By definition, the Weyl tensor of the seed geometry is given by
\begin{equation}\label{Weyl-tilde}
C^\rho{}_{\beta\mu\nu} = \frac{1}{2} 
\big(  R^\rho{}_{\beta\mu\nu} + R^\rho{}_{\nu}  g_{\beta\mu} 
+  R_{\beta\mu} \delta^\rho{}_{\nu} \big) 
+ \frac{1}{6}  R {\delta}^\rho{}_{\mu}  g_{\beta\nu} \, - \, [\mu \leftrightarrow \nu] \,.
\end{equation}
As a first step, let us derive the relation between  $\tilde{C}_{\alpha\beta\mu\nu}$ and $C_{\alpha\beta\mu\nu}$ and split it in a suitable form. Under the disformal map~\eqref{DT}, the covariant derivative $\nabla$ of a 1-form $W=W_{\mu}{d}x^{\mu}$ induces an extra term as~\cite{Wald:1984rg,Lobo:2017bfh}
\begin{equation}\label{CovD}
{\tilde \nabla}_\nu W_\mu =
{\nabla}_\nu W_\mu - D^\rho{}_{\mu\nu} W_\rho \;, \qquad \text{with} \qquad 
D^\rho{}_{\mu\nu} \equiv \frac{1}{2} {\tilde g}^{\rho\sigma}
\big( {\nabla}_{\mu}{\tilde g}_{\sigma\nu} + {\nabla}_{\nu}{\tilde g}_{\sigma\mu} 
- {\nabla}_{\sigma}{\tilde g}_{\mu\nu} \big) \,.
\end{equation}
Note that the geometrical object $D^\rho{}_{\mu\nu}$ connects the covariant derivatives in the Jordan and Einstein frames to each other. Indeed, it is the difference between the Christoffel symbols in the Jordan and Einstein frames and, therefore, it is a tensor. Using the transformation of the covariant derivative, the disformal transformation of the Riemann tensor can be expressed as
\begin{equation}\label{Riemann-tilde}
\tilde{R}^{\rho}{}_{\beta\mu\nu} = R^\rho{}_{\beta\mu\nu} + D^\rho{}_{\beta\mu\nu} \,,
\end{equation}
where the tensor $D^\rho{}_{\beta\mu\nu}$ reads
\begin{equation}\label{D}
D^\rho{}_{\beta\mu\nu} \equiv \nabla_\mu D^\rho{}_{\nu\beta} - \nabla_\nu D^\rho{}_{\mu\beta}
+ D^\rho{}_{\mu\sigma} D^\sigma{}_{\nu\beta} - D^\rho{}_{\nu\sigma} D^\sigma{}_{\mu\beta} \,.
\end{equation}
Contracting the l.h.s.\ of the Eq.~\eqref{Riemann-tilde}, we can easily obtain relations between the Ricci tensor and Ricci scalar in the Jordan and Einstein frames which read
\begin{equation}\label{Ricci-tilde}
\tilde{R}_{\mu\nu} = R_{\mu\nu}+ D_{\mu\nu} \,, \hspace{1cm}
{\tilde R} = \frac{1}{A} (R+D) - \frac{B}{A} \frac{(R_{\alpha\beta} + D_{\alpha\beta}) V^\alpha V^\beta}{A+BY} \,, 
\end{equation}
where $D_{\mu\nu} \equiv D^\rho{}_{\mu\rho\nu}$ and $D \equiv g^{\alpha\beta}D_{\alpha\beta}$. With these different pieces, it is straightforward to build up the Weyl tensor of the new disformed solution $(\M, \tilde{g})$ which can be decomposed as the sum of three different pieces given by
\begin{equation}\label{Weyl-JE-Sym0}
\tilde{C}_{\alpha\beta\mu\nu} = A \Big( C_{\alpha\beta\mu\nu} + D^T_{\alpha\beta\mu\nu}
+ \frac{B}{A} Z_{\alpha\beta\mu\nu} \Big) \,.
\end{equation}
The tensor $D^T_{\alpha\beta\mu\nu} = g_{\alpha \rho}D^{T\,\rho}{}_{\beta\mu\nu}$, where $D^{T\,\rho}{}_{\beta\mu\nu}$ is the Weyl part of the tensor \eqref{D} given by 
\begin{equation}\label{D-Weyl}
D^{T\,\rho}{}_{\beta\mu\nu} \equiv \frac{1}{2} 
\big( {D}^\rho{}_{\beta\mu\nu} + {D}^\rho{}_{\nu} {g}_{\beta\mu} 
+ { D}_{\beta\mu} \delta^\rho{}_{\nu} \big) 
+ \frac{1}{6} { D} {\delta}^\rho{}_{\mu} { g}_{\beta\nu} \, - \, [\mu \leftrightarrow \nu] \,.
\end{equation}
The tensor $Z_{\alpha\beta\mu\nu}$ is defined as
\begin{equation}\label{Z}
Z_{\alpha\beta\mu\nu} \equiv E_{\alpha\beta\mu\nu} + V_\alpha
\Big( C_{\rho\beta\mu\nu} + D^T_{\rho\beta\mu\nu} 
+ \frac{B}{A} E_{\rho\beta\mu\nu} \Big) V^\rho \,,
\end{equation} 
with
\begin{eqnarray}\label{E}
E^\rho{}_{\beta\mu\nu} &\equiv& \frac{1}{2} \bigg\{
(R^\rho{}_{\nu} + D^\rho{}_{\nu}) V_\beta V_\mu
- \frac{(R_{\sigma\nu}+D_{\sigma\nu})V^\sigma V^\rho}{A+BY} 
( A g_{\beta\mu} + B V_\beta V_\mu ) - \, [\mu \leftrightarrow \nu] \, 
\bigg\} 
\\ \nonumber
&+&\frac{1}{6} \bigg\{
 (R+D) \delta^\rho{}_\mu V_\beta V_\nu 
- \frac{(R_{\sigma\alpha}+D_{\sigma\alpha})V^\sigma V^\alpha}{A+BY} \delta^\rho{}_\mu
( A g_{\beta\nu} + B V_\beta V_\nu ) - \, [\mu \leftrightarrow \nu] \, 
\bigg\} \,,
\end{eqnarray}
which satisfies $g^{\alpha\beta}E_{\alpha\mu\beta\nu} = 0$ and $g^{\mu\nu}E_{\alpha\mu\beta\nu} \neq 0$. 

Let us now comment on the symmetries of the new terms. They will be crucial when decomposing the disformed Weyl tensor in term of the Weyl-type and Ricci-type scalars. Contrary to the two first pieces entering in the expression (\ref{Weyl-JE-Sym0}), the last tensor $Z_{\alpha\beta\mu\nu}$ is not traceless so that we can define
\begin{eqnarray}\label{Z-ab}
Z_{\mu\nu} &\equiv& g^{\alpha\beta} Z_{\alpha\mu\beta\nu} 
= \Big( C_{\alpha\mu\beta\nu} + D^T_{\alpha\mu\beta\nu} 
+ \frac{B}{A} E_{\alpha\mu\beta\nu} \Big) V^\alpha V^\beta \,, \nonumber \\
Z &\equiv& g^{\alpha\beta}Z_{\alpha\beta} 
= \frac{BY}{A+BY} (R_{\alpha\beta} + D_{\alpha\beta}) V^\alpha V^\beta \,.
\end{eqnarray} 
Since the tensor $Z_{\alpha\beta\mu\nu}$ shares the same properties as the Riemann tensor, this suggests to decompose it as
\begin{eqnarray}\label{Z-decom}
 Z_{\alpha\beta\mu\nu}& =& Z^T_{\alpha\beta\mu\nu}  +Z^S_{\alpha\beta\mu\nu} \,,
 \\ \label{Z-S}
Z^S_{\alpha\beta\mu\nu} & \equiv&
-\frac{1}{2} \big(Z_{\alpha\nu} g_{\beta\mu} + Z_{\beta\mu} g_{\alpha\nu}
- Z_{\alpha\mu} g_{\beta\nu} - Z_{\beta\nu} g_{\alpha\mu} \big) 
- \frac{1}{6} Z (g_{\alpha\mu} g_{\beta\nu} - g_{\alpha\nu} g_{\beta\mu}) \,,
\end{eqnarray}
where  $Z^T_{\alpha\beta\mu\nu} $ is the trace-free Weyl part, i.e.\ satisfying $g^{\alpha\beta}Z^T_{\alpha\mu\beta\nu}= 0$, while the trace of the second part reads $g^{\alpha\beta}Z^S_{\alpha\mu\beta\nu} = Z_{\mu\nu}$, and $g^{\alpha\beta}g^{\mu\nu}Z^S_{\alpha\mu\beta\nu}=Z$. Using this splitting, we then define a new tensor $B_{\alpha\beta\mu\nu}$ which captures the trace-free contribution to the two last pieces of (\ref{Weyl-JE-Sym0}) such that
\begin{eqnarray}\label{B}
&&B_{\alpha\beta\mu\nu} \equiv D^T_{\alpha\beta\mu\nu}
+ \frac{B}{A} Z^T_{\alpha\beta\mu\nu} \,.
\end{eqnarray}
By construction, this tensor has the same properties as the Weyl tensor and vanishes when $B=0$, namely for a pure conformal transformation. At the end of the day, the Weyl tensor in the Jordan and Einstein frames are related by
\begin{eqnarray}\label{Weyl-JE-Sym}
\tilde{C}_{\alpha\beta\mu\nu} = A \Big( C_{\alpha\beta\mu\nu} + B_{\alpha\beta\mu\nu} 
+ \frac{B}{A} Z^S_{\alpha\beta\mu\nu} \Big) \,.
\end{eqnarray}
This last expression allows one to capture the main differences between conformal and disformal transformations. It is interesting to notice that the Weyl tensor $\tilde{C}_{\alpha\beta\mu\nu}$ is no longer trace-free with respect to the untilde inverse metric $g^{\alpha\mu}$ when implementing a disformal transformation, i.e.\ when $B\neq0$, unlike the case with $B=0$.\footnote{The Weyl tensor with one upper index, which is invariant under conformal transformation can be found as $\tilde{C}^\rho{}_{\beta\mu\nu} = C^\rho{}_{\beta\mu\nu} + D^{T\,\rho}{}_{\beta\mu\nu} 
+ \frac{B}{A} E^{\rho}{}_{\beta\mu\nu} \,.$} The Weyl tensor inherits a non-traceless part with respect to the Einstein frame metric $g_{\mu\nu}$ given by $Z^S_{\alpha\beta\mu\nu}$. This difference from the conformal case shows up since the metrics in the Einstein and Jordan frames are no longer proportional to each other in the case of disformal transformation. This point will turn out to be important when decomposing this tensor on a local rest frame to build the disformed Weyl scalars. Before presenting their construction, let us nevertheless point that the Weyl tensor $\tilde{C}_{\alpha\beta\mu\nu}$ satisfies obviously the standard symmetry properties. In particular,  $\tilde{C}_{\alpha\beta\mu\nu}$ is traceless with respect to the disformal metric $\tilde{g}_{\alpha\beta}$, as should be. 

\subsection{Disformed Weyl scalars}

Following the standard approach to the Petrov classification, which amounts at identifying the number of independent PNDs of a given spacetime geometry, as well as their multiplicity, we now turn to the decomposition of the Weyl tensor. The Weyl scalars are defined by~\cite{Chandrasekhar:1985kt,Stephani:2003tm}
\begin{eqnarray}\label{WeylS-tilde-0}
\tilde{\bf \Psi}_0 & =& \tilde{C}_{\alpha\beta\mu\nu} \tilde{l}^{\alpha} {\tilde m}^{\beta} \tilde{l}^{\mu} {\tilde m}^{\nu} \,,\\
\tilde{\bf \Psi}_1 & =& \tilde{C}_{\alpha\beta\mu\nu} {\tilde l}^{\alpha} {\tilde k}^{\beta} {\tilde l}^{\mu} {\tilde m}^{\nu}  \,, \\ 
\tilde{\bf \Psi}_2 & =& {\tilde C}_{\alpha\beta\mu\nu} {\tilde l}^{\alpha} {\tilde m}^{\beta} \tilde{\bar m}^{\mu} {\tilde k}^{\nu}  \,, \\ 
\tilde{\bf \Psi}_3 & =& {\tilde C}_{\alpha\beta\mu\nu} {\tilde k}^{\alpha} {\tilde l}^{\beta}  {\tilde k}^{\mu} \tilde{\bar m}^{\nu} \,, \\ 
\tilde{\bf \Psi}_4 & =& {\tilde C}_{\alpha\beta\mu\nu} {\tilde k}^{\alpha} \tilde{\bar m}^{\beta} {\tilde k}^{\mu} \tilde{\bar m}^{\nu} \,. \label{WeylS-tilde-4}
\end{eqnarray}
Our goal now is to express these Weyl scalars $\tilde{\bf \Psi}_I$, associated to the disformed metric $\tilde{g}_{\mu\nu}$, in terms of the Weyl scalars ${\bf \Psi}_I$ associated to the seed metric $g_{\mu\nu}$. This will allow us to provide closed expressions which capture how the Petrov type of a given spacetime geometry changes under a disformal map. To do so, we rewrite the Weyl scalars in Jordan frame~\eqref{WeylS-tilde-0}-\eqref{WeylS-tilde-4}  in terms of the null tetrad in the Jordan frame. Let us first introduce the notation
\begin{eqnarray}\label{WeylS-tilde-T}
\tilde{\bf \Psi}_0 = \tilde{C}_{0202} \,, \hspace{.5cm}
\tilde{\bf \Psi}_1 = \tilde{C}_{0102} \,, \hspace{.5cm}
\tilde{\bf \Psi}_2 = {\tilde C}_{0231} \, \hspace{.5cm}
\tilde{\bf \Psi}_3 = {\tilde C}_{1013} \,, \hspace{.5cm}
\tilde{\bf \Psi}_4 = {\tilde C}_{1313}  \,,
\end{eqnarray}
where 
\begin{equation}\label{C-tilde}
\tilde{C}_{abcd} \equiv \tilde{C}_{\alpha\beta\mu\nu} 
\tilde{\theta}^{\alpha}{}_a\tilde{\theta}^{\beta}{}_b \tilde{\theta}^{\mu}{}_c \tilde{\theta}^{\nu}{}_d \,,
\end{equation}
are tetrad components of the Weyl tensor in Jordan frame with respect to the tetrads $\tilde{\theta}^{\nu}{}_a$. As it is clear from the above relations, the Weyl complex scalars are some special tetrad components of the Weyl tensor $C_{abcd}$ which describe ten independent components of the Weyl tensor. The other tetrad components of the Weyl tensor are not independent (see Eq.~\eqref{Weyl-components} in the appendix \ref{app-tetrad-rep}). Substituting Eq.~\eqref{Weyl-JE-Sym} in Eq.~\eqref{C-tilde}, we find
\begin{eqnarray}\label{C-tilde-trans0}
{\tilde C}_{efgh} = \frac{A}{4} \Big( C_{abcd} + B_{abcd} + \frac{B}{A} Z^S_{abcd} \Big) 
T^{ab}{}_{ef}T^{cd}{}_{gh} \,,
\end{eqnarray}
in which $C_{abcd} $, $B_{abcd} $, $Z^S_{abcd}$ are defined in the same manner as Eq.~\eqref{C-tilde} correspondingly. We have introduced the compact notation
\begin{eqnarray}\label{Tabcd}
T^{ab}{}_{ef} \equiv T^a{}_e T^b{}_f - T^a{}_f T^b{}_e
= \frac{1}{A} \big( \delta^a{}_e \delta^b{}_f - \delta^a{}_f \delta^b{}_e \big) 
- \frac{\beta}{A} V^{ab}{}_{ef} \,,
\end{eqnarray}
where we have used Eq.~\eqref{J-inv} in the second step. The last term is defined as
\begin{eqnarray}\label{Qabcd}
V^{ab}{}_{ef} \equiv
\delta^a{}_e V^b V_f + \delta^b{}_f V^a V_e - \delta^a{}_f V^b V_e - \delta^b{}_e V^a V_f \,.
\end{eqnarray}
Note that $T^{ab}{}_{ef}$ and $V^{ab}{}_{ef}$ are totally antisymmetric in both upper and lower pairs of indices. Moreover, $T^{ab}{}_{ef}$ is linear in $\beta$ as the quadratic parts cancel each other. Substituting Eq.~\eqref{Tabcd} in Eq.~\eqref{C-tilde-trans0} we find
\begin{eqnarray}\label{C-tilde-trans-prim}
{\tilde C}_{efgh} &=& \frac{1}{A} \Big( C_{efgh} + B_{efgh} + \frac{B}{A} Z^S_{efgh} \Big) \nonumber \\ 
&-& \frac{\beta}{2A} \Big\{ \Big( C_{efab} + B_{efab} + \frac{B}{A} Z^S_{efab} \Big) V^{ab}{}_{gh}
+ \Big( C_{ghab} + B_{ghab} + \frac{B}{A} Z^S_{ghab} \Big) V^{ab}{}_{ef} \Big\}
\nonumber \\
&+&  \frac{\beta^2}{4A} \Big( C_{abcd} + B_{abcd} + \frac{B}{A} Z^S_{abcd} \Big)
V^{ab}{}_{ef} V^{cd}{}_{gh} \,.
\end{eqnarray}
Now, the question is how many scalar quantities do we need to fully encodes the information encapsulated in the r.h.s.\ of the above equation. In the Einstein frame, the symmetry of the Weyl tensor allows one to use only the standard five complex Weyl scalars ${\bf \Psi}_I$ to express all components of ${C}_{abcd}$ as it is done in Eq.~\eqref{WeylS-T-a} in the appendix \ref{app-tetrad-rep}. By construction, the tensor $B_{\alpha\beta\mu\nu}$ has the same symmetry as the Weyl tensor and we only need five more complex Weyl-type scalar, which we denote ${\mathbf \Delta}_I$ ($I=0,\cdots,4$), in order to decompose it. They read
\begin{eqnarray}\label{Delta}
{\mathbf \Delta}_0 = B_{0202} \,, \hspace{.5cm}
{\mathbf \Delta}_1 = B_{0102} \,, \hspace{.5cm}
{\mathbf \Delta}_2 = B_{0231} \,, \hspace{.5cm}
{\mathbf \Delta}_3 = B_{1013} \,, \hspace{.5cm}
{\mathbf \Delta}_4 = B_{1313} \,.
\end{eqnarray}
The decomposition of the remaining trace part $Z^S_{abcd}$ is more subtle. Indeed, as pointed in the previous section, it does not share the same symmetry as the Weyl tensor and one therefore needs additional scalars to fully capture the information encoded in $Z^S_{abcd}$. From the definition Eq.~\eqref{Z-S} we find
\begin{eqnarray}\label{Zij}
Z^S_{abcd} =  
-\frac{1}{2} \big(Z_{ad} \eta_{bc} + Z_{bc} \eta_{ad}
- Z_{ac} \eta_{bd} - Z_{bd} \eta_{ac} \big) 
- \frac{1}{6} Z (\eta_{ac} \eta_{bd} - \eta_{ad} \eta_{bc}) \,,
\end{eqnarray}
which shows that $Z^S_{abcd}$ is completely expressed in terms of its Ricci part $Z_{ab}$. Therefore, it can be completely expressed in terms of the four real and three complex scalars given by
\begin{eqnarray}\label{Pi}
&&{\bf{\Pi}}_{00} = \frac{1}{2} Z_{00} \,, \hspace{.5cm}
{\bf{\Pi}}_{11} = \frac{1}{4} (Z_{01}+Z_{23}) \,, \hspace{.5cm}
{\bf{\Pi}}_{22} = \frac{1}{2} Z_{11} \,, \hspace{.5cm}
 {\bf{\Lambda^S}} = - \frac{1}{24} Z = \frac{1}{12} (Z_{01} - Z_{23}) \,, \nonumber \\
&&{\bf{\Pi}}_{01} = \frac{1}{2} Z_{02} \,, \hspace{1.5cm}
{\bf{\Pi}}_{02} = \frac{1}{2} Z_{22} \,, \hspace{1.5cm}
{\bf{\Pi}}_{12} = \frac{1}{2} Z_{12} \,.
\end{eqnarray}
These additional Ricci-type scalars, i.e.\ ${\bf \Delta}_I$ and ${\bf \Pi}_{IJ}$ and ${\bf \Lambda}^S$, allows one to write close relations between the Weyl scalars ${\bf \tilde{\Psi}}_I$  and  ${\bf \Psi}_I$ which can be compactly written as 
\begin{eqnarray}\label{C-tilde-trans}
\tilde{\bf \Psi}_I = \tilde{\bf \Psi}_I({\bf \Psi}_I,{\bf \Delta}_I,{\bf \Pi}_{IJ})  \,.
\end{eqnarray}
As a concrete example, the first disformed Weyl scalar ${\bf \tilde{\Psi}}_0$ decomposes as
\begin{eqnarray}\label{Psi0-mail}
A \tilde{\boldsymbol \Psi}_0 &=&  \gamma^2 ( {\boldsymbol \Psi}_0 + {\boldsymbol \Delta}_0 )
+ \beta^2 \big[ ( {\boldsymbol \Psi}^*_0 + {\boldsymbol \Delta}^*_0 ) (V_2V_2)^2 
+ ( {\boldsymbol \Psi}^*_4 + {\boldsymbol \Delta}^*_4 ) (V^1V^1)^2 \big]
\nonumber \\ 
&+& 2 \beta^2 V^1 V_2 \big[ 2 ( {\boldsymbol \Psi}^*_1 + {\boldsymbol \Delta}^*_1 ) V_2V_2 
+ 3 ( {\boldsymbol \Psi}^*_2 + {\boldsymbol \Delta}^*_2 ) V^1V_2 
+ 2 ( {\boldsymbol \Psi}^*_3 + {\boldsymbol \Delta}^*_3 ) V^1V^1 \big] 
\nonumber \\ 
&-& 2 \frac{B}{A} \beta \gamma \big[ 2 {\boldsymbol \Pi}_{01} V^1V_2 
+ {\boldsymbol \Pi}_{02} V^1 V^1 + {\boldsymbol \Pi}_{00} V_2 V_2 \big]
\,.
\end{eqnarray}
The explicit form of disformed Weyl scalars being quite involved, we provide their decomposition in the Appendix~\ref{app-tetrad-rep} which are given by the relations~\eqref{Psi0},~\eqref{Psi1},~\eqref{Psi2},~\eqref{Psi3}, and~\eqref{Psi4}. This provides the main result of this work.

Having the decomposition of all Weyl scalars ~\eqref{C-tilde-trans} in hand, we can study the Petrov classification after disformal transformation. The classification is performed using the following Lorentz-invariant quantities
\begin{eqnarray}\label{Petrov-quantities}
&&{\tilde I} \equiv \tilde{\bf \Psi}_0 \tilde{\bf \Psi}_4 - 4 \tilde{\bf \Psi}_1 \tilde{\bf \Psi}_3 
+ 3 \tilde{\bf \Psi}_2^2 \,, \hspace{1cm}
{\tilde J} \equiv 
\begin{vmatrix}
\tilde{\bf \Psi}_4 & \tilde{\bf \Psi}_3 & \tilde{\bf \Psi}_2 \\ 
\tilde{\bf \Psi}_3 & \tilde{\bf \Psi}_2 & \tilde{\bf \Psi}_1 \\ 
\tilde{\bf \Psi}_2 & \tilde{\bf \Psi}_1 & \tilde{\bf \Psi}_0
\end{vmatrix} \,, \hspace{1cm} 
{\tilde D} \equiv {\tilde I}^3 - 27 {\tilde J}^2 \,,
\\ \nonumber
&& {\tilde K} \equiv \tilde{\bf \Psi}_4^2 \tilde{\bf \Psi}_1 
- 3 \tilde{\bf \Psi}_4 \tilde{\bf \Psi}_3 \tilde{\bf \Psi}_2 + 2 \tilde{\bf \Psi}_3^3 \,, \hspace{1cm}
{\tilde L} \equiv \tilde{\bf \Psi}_4 \tilde{\bf \Psi}_2 - \tilde{\bf \Psi}_3^2 \,, \hspace{1cm}
{\tilde N} \equiv 12 {\tilde L}^2 - \tilde{\bf \Psi}_4^2 {\tilde I} \,.
\end{eqnarray} 
Depending on the values of these scalars, one can directly determine in a coordinate-independent and Lorentz-invariant manner the Petrov type of a given geometry. The classification is summarized in Table \ref{tab1}. Physically, the different types can be understood as follows in terms of their PNDs. A given geometry possesses four distinct PNDs at each point when it is type I, three PNDs with one repeated for type II, one triply repeated for type III, two PNDs both repeated for type D and finally only one PND but quadruply repeated for type N.
\begin{table}
	\centering
	\begin{tabular}{ |p{1cm}|p{7cm}|  }
		\hline
		\multicolumn{2}{|c|}{\bf Petrov Classification} \\
		\hline 
		\hfil Type & \hfil Conditions \\
		\hline
		\hfil O & $\tilde{\bf \Psi}_0=\tilde{\bf \Psi}_1=\tilde{\bf \Psi}_2=\tilde{\bf \Psi}_3=\tilde{\bf \Psi}_4=0$ \\
		\hfil I & $\tilde{ D}\neq0$ \\
		\hfil II & $\tilde{ D} =0$, ${\tilde I}\neq0$, ${\tilde J}\neq0$, ${\tilde K}\neq0$, ${\tilde N}\neq0$ \\
		\hfil III & ${\tilde D} =0$, ${\tilde I} = {\tilde J} =0$, ${\tilde K}\neq0$, ${\tilde L}\neq0$ \\
		\hfil N & ${\tilde D} =0$, ${\tilde I}={\tilde J}={\tilde K}={\tilde L}=0$ \\
		\hfil D & ${\tilde D} =0$, ${\tilde I}\neq0$, ${\tilde J}\neq0$, ${\tilde K}={\tilde N}=0$\\
		\hline
	\end{tabular}
	\newline
	\caption{The left column shows the Petrov types and the right column shows the corresponding desired conditions. Quantities ${\tilde I}, {\tilde J}, {\tilde D}, {\tilde K}, {\tilde L}$, and ${\tilde N}$ are defined in Eq.~\eqref{Petrov-quantities} in terms of the Weyl scalars $\tilde{\bf \Psi}_I$.}\label{tab1}
\end{table}

The results that we found in this section show that Petrov type of a given spacetime geometry can change after performing a disformal transformation. We will explicitly confirm this fact in the next section by applying our general formalism to some concrete examples. Moreover, having disformed Weyl scalars~\eqref{C-tilde-trans} in hand, we can look for conditions under which Petrov types are invariant. Therefore, one can extract general results from our analysis which are useful especially when constructing a given exact solution through a disformal transformation.

Before closing this subsection, let us consider the case of conformal map by means of our formalism. In this case, $B=0$ such that $B_{abcd} = 0$. From the results~\eqref{Psi0}-\eqref{Psi4}, it follows that $\tilde{\bf \Psi}_I = A^{-1} {\bf \Psi}_I$. From the definitions~\eqref{Petrov-quantities}, we find ${\tilde I} = A^{-2} I$, ${\tilde J} = A^{-3} J$, ${\tilde D} = A^{-6} D$, ${\tilde K} = A^{-3} K$, ${\tilde L} = A^{-2} L$, and ${\tilde N} = A^{-4} N$ where all quantities without tilde correspond to the definitions~\eqref{Petrov-quantities} replacing $\tilde{\bf \Psi}_I$ with ${\bf \Psi}$. Therefore, the Petrov type of a spacetime geometry will not change under a conformal transformation~\cite{Ajith:2020ydz}, as expected.

\subsection{Simplified disformal transformation}\label{subsec-pure-disformal}

\label{secgencond}

In this subsection, we use the results obtained in this section and analyze them in a simplified setting. As the effect of conformal transformation is already well understood, we focus on the pure disformal case with $A=1$ and constant small disformal factor $B=B_0\ll1$. Indeed, considering only constant value of the conformal factor $A=A_0$, without loss of generality $A_0=1$ can be realized through redefinition of metric. The disformal transformation~\eqref{DT} then takes the form
\begin{equation}\label{DT-S}
{\tilde g}_{\mu\nu} = {g}_{\mu\nu} + B_0\, V_\mu V_\nu \,,
\end{equation}
The above transformation captures all essential features of the disformal map while the analysis becomes significantly simpler. Moreover, an interesting application of the disformal mapping is to consider a transformation for which $ B_0\ll1$. It corresponds to small deviations form GR and provide thus an interesting regime relevant for the comparison with observations. Starting from this setup, we shall now provide simplified expressions for the relations between the Weyl scalars before and after the disformal transformations up to first order in $B_0$.

For the choice $A=1$ and $B=B_0\ll1$, Eq.~\eqref{beta} implies that $\beta \approx \frac{B_0}{2}$. The null tetrad in Jordan frame, given by Eq.~\eqref{kl-disform}, simplifies to
\begin{eqnarray}\label{kl-disform-exp}
&&{\tilde l}^\mu \approx 
{l}^\mu{} + \frac{B_0}{2} V^1 V^\mu \,, 
\hspace{1.5cm}
{\tilde k}^\mu \approx 
{k}^\mu{} - \frac{B_0}{2} V_1 V^{\mu}\,,
\nonumber \\
&&{\tilde m}^\mu \approx
{m}^\mu{} - \frac{B_0}{2} V_2 V^\mu \,, 
\hspace{1cm}
{\tilde {\bar m}}^\mu \approx
{{\bar m}}^\mu{} - \frac{B_0}{2} V^2 V^\mu \,.
\end{eqnarray}
Using these expressions, and expanding for $B=B_0\ll1$, the results~\eqref{Psi0},~\eqref{Psi1},~\eqref{Psi2},~\eqref{Psi3}, and~\eqref{Psi4} simplify to
\begin{eqnarray}\label{Psi0-ex}
\tilde{{\boldsymbol \Psi}}_0 & \approx &
{\boldsymbol \Psi}_0 \big( 1 - B_0 ( V_1V^1+V_2V^2 ) \big) +  {\boldsymbol \Delta}_0 \,,
\\[7pt]
\label{Psi1-ex}
\tilde{\boldsymbol \Psi}_1 & \approx &
{\boldsymbol \Psi}_1 \Big( 1 - \frac{B_0}{2} ( V_1V^1+V_2V^2 ) \Big) + {\boldsymbol \Delta}_1
\nonumber \\
&-& \frac{B_0}{2} \Big[ {\boldsymbol \Psi}_0 V_1 V^2
+ {\boldsymbol \Psi}^*_1 V_2 V_2
+ (2 {\boldsymbol \Psi}^*_2 - {\boldsymbol \Psi}_2 ) V^1 V_2 
+ {\boldsymbol \Psi}^*_3 V^1 V^1 
- 2 {\boldsymbol \Pi}_{01} \Big] \,,
\end{eqnarray}

\begin{eqnarray}\label{Psi2-ex}
\tilde{\boldsymbol \Psi}_2 & \approx &
{\boldsymbol \Psi}_2 \big( 1 - B_0 ( V_1V^1+V_2V^2 ) \big) + {\boldsymbol \Delta}_2
- 2B_0 {\boldsymbol \Lambda}^S \,,
\\[7pt]
\label{Psi3-ex}
\tilde{\boldsymbol \Psi}_3 & \approx &
{\boldsymbol \Psi}_3 \Big( 1 - \frac{B_0}{2} ( V_1V^1+V_2V^2 ) \Big) + {\boldsymbol \Delta}_3 
\nonumber \\ 
&-& \frac{B_0}{2} \Big[ {\boldsymbol \Psi}^*_1 V_1 V_1 
- (2 {\boldsymbol \Psi}^*_2 - {\boldsymbol \Psi}_2 ) V_1 V^2 - {\boldsymbol \Psi}_4 V^1 V_2
+ {\boldsymbol \Psi}^*_3 V^2 V^2 - 2 {\boldsymbol \Pi}^*_{12} \Big] \,,
\\[7pt]
\label{Psi4-ex}
\tilde{\boldsymbol \Psi}_4 & \approx &
{\boldsymbol \Psi}_4 \Big( 1 - B_0 ( V_1V^1+V_2V^2 ) \Big) + {\boldsymbol \Delta}_4 \,.
\end{eqnarray}
From the above results, it is more clear that Petrov types generally change under disformal transformation. For instance, suppose that we have a type O solution in Jordan frame so that all $\tilde{\boldsymbol \Psi}_I = 0$. We thus have five equations for five complex variables ${\boldsymbol \Psi}_I$ in the Einstein frame which give non-vanishing solutions for ${\boldsymbol \Psi}_I$ in general. Therefore a type O metric in Jordan frame will be no longer type O in the Einstein frame in general.

Having presented the simplified expressions for the disformed Weyl scalars for the disformal map Eq.~\eqref{DT-S}, we can now apply our general results to investigate concrete known exact solutions. This is the subject of the next section.

\section{Applications to concrete examples}\label{sec4}

In this section, we apply our general setup to several relevant solutions of the modified gravity theories to illustrate how the Petrov type of a given seed solution changes after performing a disformal transformation. We consider three different seed configurations: i) a FLRW cosmology, ii) a stealth spherically symmetric black hole and finally iii) the case of a stealth Kerr geometry. For the sake of simplicity, we perform all analysis for a pure disformal map with $B_0\ll 1$ presented in subsection \ref{subsec-pure-disformal}.

\subsection{Scalar field cosmology}

Let us first consider a simple example of the FLRW cosmology. The line element for the seed solution is given by the spatially curved FLRW background in spherical coordinates $(t,r,\theta,\varphi)$ given by
\begin{equation}
\label{ds2-FLRW}
ds^2_{\rm FLRW} = - dt^2 + a(t)^2 \bigg[ 
\frac{dr^2}{1-kr^2} + r^2 d\Omega^2 
\bigg] \,.
\end{equation}
In the line element Eq.~\eqref{ds2-FLRW}, $a(t)$ is the scalar factor, $d\Omega^2 = d\theta^2 + \sin^2\theta d\varphi^2$ is the metric of the unit sphere, and $k$ is the normalized constant curvature of the spatial metric which takes the values $k=0,1,-1$ for flat, spherical, and hyperbolic 3-dimensional spatial sections respectively. Consider the following PNDs \cite{GomezLopez:2017kcw}
\begin{align}\label{null-vectors-FLRW}
l^\mu & = \frac{1}{\sqrt{2}} \Big( 1 , - \frac{\sqrt{1-k r^2}}{a},0 , 0 \Big) ,\\
\hspace{.3cm}
k^\mu & = \frac{1}{\sqrt{2}} \Big( 1 ,  \frac{\sqrt{1-k r^2}}{a}, 0 , 0 \Big) , \label{null-vectors-FLRW-k}\\
\hspace{.3cm}
m^\mu & = \frac{1}{\sqrt{2}r} \Big( 0, 0 , - \frac{1}{a}, \frac{i}{\sin\theta} \Big) ,\label{null-vectors-FLRW-m}
\end{align}
which satisfy the desired conditions~\eqref{null} as well as the vanishing of other contractions. Using the above PNDs, it is easy to confirm that ${\mathbf \Psi}_I=0$ for all $I$ and the seed FLRW solution~\eqref{ds2-FLRW} is type O as it is well known. Indeed, this is clear since the metric~\eqref{ds2-FLRW} is conformally flat and therefore the corresponding Weyl tensor vanishes.  

In order to study the effect of a disformal transformation on such cosmological background, one needs to preserve homogeneity and isotropy. The natural choice for the disformal vector is therefore $V_\mu= \partial_\mu \phi$ in which $\phi$ is a scalar field with time-dependent homogeneous vev $\phi(t)$. The vector field then takes the simple form
\begin{equation}\label{V-FLRW}
V_\mu = (\dot{\phi},0,0,0) \,,
\end{equation}
where a dot denotes derivative with respect to the time $t$. The disformed metric takes the following form
\begin{align}\label{ds2-FLRW-dis}
{\widetilde ds}^2_{\rm{FLRW}} & = { ds}^2_{\rm{FLRW}} + B_0 \dot{\phi}^2 dt^2 \\
& = - (1-B_0 \dot{\phi}^2) dt^2 + a(t)^2 \bigg[ 
\frac{dr^2}{1-kr^2} + r^2 d\Omega^2 
\bigg] \,.
\end{align}
The effects of disformal transformation in this case turns out to be quite trivial as we can define a new time coordinate $t'=\int (1-B_0 \dot{\phi}^2)^{1/2} dt$ in terms of which the metric takes the standard FLRW form. One can find the new PNDs and Weyl scalars after disformal transformation by substituting~\eqref{null-vectors-FLRW}-\eqref{V-FLRW} in~\eqref{kl-disform-exp} and~\eqref{Psi0-ex}-\eqref{Psi4-ex} respectively. However, as we have shown above, they will not change after redefining the time coordinate. Therefore, the Petrov type of the metric~\eqref{ds2-FLRW} will not change and remains type O.

\subsection{Disformal Schwarzschild-like solution}

We now turn to the case where the seed solution is a spherically symmetric, static geometry of type D. A vast landscape of such exact black hole solutions has been constructed for DHOST theories~\cite{Kobayashi:2014eva, Babichev:2017guv, BenAchour:2018dap, Motohashi:2018wdq, Minamitsuji:2018vuw, Bernardo:2019yxp, Motohashi:2019sen, Takahashi:2020hso, Charmousis:2019vnf}. Before discussing these solutions, it is worth pointing that such exact stealth solutions usually suffer from the strong coupling issue~\cite{Minamitsuji:2018vuw,deRham:2019gha,Motohashi:2019ymr,Khoury:2020aya,Langlois:2021aji} in the case of scalar-tensor theories, which can nevertheless be cured through the implementation of the scordatura scenario as far as $\partial_\mu\phi$ is timelike~\cite{Motohashi:2019ymr,Gorji:2020bfl}.  Exact spherically symmetric black hole solutions are also available in vector-tensor theories \cite{Cisterna:2016nwq, Chagoya:2016aar, Chagoya:2017ojn, Heisenberg:2017xda, Babichev:2017rti, Heisenberg:2017hwb, Kase:2018voo, Kase:2018owh, Minamitsuji:2021gcq} while some aspects of black holes in mimetic gravity can be found in \cite{Gorji:2019rlm, BenAchour:2017ivq, Yoshida:2018kwy}. We refer the reader to \cite{BenAchour:2019fdf} for a detail discussion on the construction of disformal spherical symmetric solutions in DHOST theories.

\subsubsection{Standard kinematics}

Consider thus a seed solution whose line element is given by
\begin{equation}\label{ds2-SS}
ds^2_{\rm SS} = - f(r) dt^2+ \frac{dr^2}{f(r)} + r^2 d\Omega^2 \,,
\end{equation}
which includes for example the form of the metric of a stealth Schwarzschild black hole. Notice that this is not the most general case, yet it provides a relevant example of the backgrounds used in modified gravity as seed.

 For such geometry, a set of PNDs is given by
\begin{eqnarray}\label{null-vectors-SS}
l^\mu & = \frac{1}{\sqrt{2}} \Big( \frac{1}{\sqrt{f}} , - \sqrt{f}, 0 , 0 \Big) , \\
\hspace{.3cm}
k^\mu & = \frac{1}{\sqrt{2}} \Big( \frac{1}{\sqrt{f}} , \sqrt{f}, 0 , 0 \Big) , \\
\hspace{.3cm}
m^\mu & = \frac{1}{\sqrt{2}r} \Big( 0 , 0 , - 1, \frac{i}{\sin\theta} \Big) ,
\end{eqnarray}
and one can check that the only non-zero Weyl scalar is
\begin{equation}\label{Weyl-SS}
{\boldsymbol \Psi}_2 = \frac{1}{6r^2} \Big( -1 + f - r f' + \frac{r^2}{2} f'' \Big) \,.
\end{equation}
The other nonzero scalars constructed by the tetrad components of the Ricci tensor which are defined in Eq.~\eqref{RicciS-T-a} are also given by
\begin{eqnarray}\label{RicciS-SS}
{\boldsymbol \Phi}_{11} = \frac{1}{4r^2} \Big( 1 - f + \frac{r^2}{2} f'' \Big) \,,
\hspace{1cm}
{\boldsymbol \Lambda} = \frac{1}{12r^2} \Big( -1 + f + 2 r f' + \frac{r^2}{2} f'' \Big) \,.
\end{eqnarray}
Therefore, using the definitions~\eqref{Petrov-quantities} and the Table \ref{tab1}, one finds as expected that the seed metric~\eqref{ds2-SS} is of Petrov type D. We can now investigate how the Petrov type changes under a disformal mapping depending on the profile of the additional field $V$.

\subsubsection{Disformed kinematics}

In order not to spoil the spherical symmetry of the seed metric~\eqref{ds2-SS}, we consider a general vector field with components depending only on the time $t$ and radial $r$ coordinates.\footnote{We have so far considered the case of a static spherically symmetric case. As for the general spherically symmetric solution (aimed to describe, e.g.\ a collapse), which can be written as $ds^2=g_{\mu\nu}\,dx^\mu dx^\nu=-f_1(t,r)^2\,dt^2+dr^2/f_2(t,r)^2+r^2\,d\Omega^2$, and for its disformed case, that is $\tilde g_{\mu\nu}=A(t,r)\,g_{\mu\nu}+B(t,r)V_\mu V_\nu $, where $V_\mu\,dx^\mu = V_t(t,r)\,dt+ V_r(t,r)\,dr$, it can be shown, by direct calculations, that they are both of Petrov type D.} Without loss of generality, the vector profile reads
\begin{equation}\label{SS-V-mu}
V_\mu = \big( V_t, V_r, 0, 0 \big) \,, \hspace{1cm} 
V^\mu = \Big( -\frac{V_t}{f}, f V_r, 0, 0 \Big) \,,
\end{equation}
where $V_t:= V_t(t,r)$ and $V_r:=V_r(t,r)$. The tetrad components of the vector field defined in Eq.~\eqref{Va} are given by 
\begin{align}\label{SS-V-a}
V_a & = \frac{1}{\sqrt{2f}} \big( V_t-f V_r , V_t + f V_r, 0, 0 \big) \,, \hspace{1cm} \\
V^a & = - \frac{1}{\sqrt{2f}} \big( V_t+ f V_r, V_t - f V_r, 0, 0 \big) \,.
\end{align}
We see that $V_3=V^2=0$ and $V^3=V_2=0$ which is equivalent to the orthogonality condition~\eqref{orthogomal-mmbar}. Thus, the disformal vector $V^\mu$, given by Eq.~\eqref{SS-V-mu}, lives on the tangent space $T_p({\cal M}/{\cal S})$ and is orthogonal to the 2-dimensional surface ${\cal S}$. 

Applying the above results to the seed metric~\eqref{ds2-SS}, we find that the spherically symmetric disformed metric takes the form
\begin{equation}\label{ds2-SS-dis}
{\widetilde ds}^2_{\rm{SS}} = {ds}^2_{\rm{SS}} + B_0 \big( V_t dt + V_r dr \big)^2 \,.
\end{equation}
Starting from the PNDs in Einstein frame Eq.~\eqref{null-vectors-SS}, we can easily find the disformed PNDs which read
\begin{align}\label{null-vectors-SS-dis}
{\tilde l}^\mu & = l^\mu - \frac{B_0}{2} 
\Big( \frac{V_t - f V_r }{\sqrt{2f}} \Big) V^\mu \,, \hspace{1cm}\\
{\tilde k}^\mu & = k^\mu - \frac{B_0}{2} 
\Big( \frac{V_t + f V_r }{\sqrt{2f}} \Big) V^\mu \,, \label{null-vectors-SS-dis-k}
\\
\hspace{1cm}
{\tilde m}^\mu  &= m^\mu \,,\label{null-vectors-SS-dis-m}
\end{align}
where the explicit forms of undisformed PNDs and the disformal vector $V^\mu$ are given by Eqs.~\eqref{null-vectors-SS} and~\eqref{SS-V-mu} respectively.
Now, let us consider the Weyl-type and Ricci-type scalars to investigate the fate of the Petrov type. From definitions~\eqref{Delta} and~\eqref{Pi}, we find the following nonzero scalars
\begin{eqnarray}\label{Dis-Scalars-SS}
&&{\boldsymbol \Delta}_2 =
\frac{B_0}{12 r^2}
\Bigg\{
\frac{V_t^2}{f} - V_t^2 - f V_r^2 
- 2 r V_r \dot{V}_t
- f^2 V_r \left(V_r-2 r V_r'\right)
- 2 r V_t \big(r V_t''-r\dot{V}_r'+\dot{V}_r-V_t' \big)
\nonumber \\ \nonumber
&& \hspace{2.3cm}
- 2 r^2 \left(V_r (\ddot{V}_r-\dot{V}_t' )-\dot{V}_r V_t'-\dot{V}_t
V_r'+\dot{V}_r^2+ V_t'^2\right)
+ \frac{r^2 f''}{2f}
\left(V_t^2-f^2 V_r^2\right)
\\ \nonumber
&& \hspace{2.3cm} - \frac{r^2 f'^2}{2f^2}
\left( V_t^2 + f^2 V_r^2\right)
+ \frac{r f'}{f} \left(f^2 V_r
\left(2 V_r-r V_r'\right)+r V_t \left(\dot{V}_r+V_t'\right)+r V_r
\dot{V}_t\right) \Bigg\} \,,
\\
&& {\boldsymbol \Pi}_{00} = \frac{{\boldsymbol \Psi}_2}{2f} \big( V_t - f V_r \big)^2 \,, 
\hspace{.5cm}
{\boldsymbol \Pi}_{11} = - \frac{{\boldsymbol \Psi}_2}{2f} \big( V_t^2 - f^2 V_r^2 \big) \,,
\hspace{.5cm}
{\boldsymbol \Pi}_{22} = \frac{{\boldsymbol \Psi}_2}{2f} \big( V_t + f V_r \big)^2 \,,
\end{eqnarray}
where dot and prime denote derivative with respect to the coordinates $t$ and $r$ respectively. In the case of spherically symmetric metric Eq.~\eqref{ds2-SS}, using Eqs.~\eqref{Weyl-SS},~\eqref{RicciS-SS}, ~\eqref{SS-V-a}, and~\eqref{Dis-Scalars-SS}, Eqs.~\eqref{Psi0-ex}-\eqref{Psi4-ex} simplify to
\begin{eqnarray}\label{Psi-SS}
\tilde{\boldsymbol \Psi}_0 = \tilde{\boldsymbol \Psi}_1 
= \tilde{\boldsymbol \Psi}_3 = \tilde{\boldsymbol \Psi}_4 = 0 \,, \hspace{1cm}
\tilde{\boldsymbol \Psi}_2 = {\boldsymbol \Psi}_2 
\Big( 1 + \frac{B_0}{2f} \big(V_t^2 - f^2 V_r^2 \big) \Big) + {\boldsymbol \Delta}_2 \,,
\end{eqnarray}
where the explicit form of ${\boldsymbol \Delta}_2$ is given by~\eqref{Dis-Scalars-SS}. Looking at the Table \ref{tab1}, the result~\eqref{Psi-SS} shows that the disformed metric~\eqref{ds2-SS-dis} is again of Petrov type D. This is consistent with the fact that the disformed metric~\eqref{ds2-SS-dis} is spherically symmetric. Indeed, it is well known that all spherically symmetric spacetimes are of Petrov type D (or O). 

We now turn to the axisymmetric case.

\subsection{Disformal Kerr-like solution}

While spherically symmetric solutions are of interest to understand the description of compact object, rotating configurations are the ones required to confront modified gravity theories to current astrophysical observations. In the DHOST context, a stealth Kerr-(A)dS solution dressed with a linear time-dependent scalar profile has been derived in~\cite{Charmousis:2019vnf} while general conditions for the existence of such stealth configurations have been addressed in~\cite{Takahashi:2020hso}. More recently, such seed was used to construct the first non-stealth Kerr-like black hole solution for a large family of DHOST theories~\cite{BenAchour:2020fgy,Anson:2020trg}. See~\cite{ Long:2020wqj, Anson:2021yli} for recent investigations on its properties. In the following, we derive its Petrov type and show that the new solution changes the Petrov type.   See \cite{Chagoya:2016aar,Babichev:2017rti,Cisterna:2016nwq,Chagoya:2017ojn,Ajith:2020ydz} for stealth rotating black hole in vector-tensor theories

\subsubsection{Standard kinematics}
We consider the seed metric to be the stealth Kerr solution given by
\begin{equation}\label{ds2-Kerr}
ds^2_{\rm Kerr} = -\frac{\Delta}{\rho^2} (dt - a \sin^2\theta d\varphi)^2 + \frac{\rho^2}{\Delta} dr^2 + \rho^2 d\theta^2 + \frac{\sin^2\theta}{\rho^2} 
\big( a dt - (r^2+a^2) d\varphi \big)^2 \,,
\end{equation}
where $M$ and $a$ are mass and angular momentum per unit mass and we have defined
\begin{equation}
\Delta \equiv r^2 - 2 Mr + a^2 \,, \hspace{1cm} \rho^2 \equiv r^2 + a^2 \cos^2\theta \,.
\end{equation} 
Following Ref.~\cite{Chandrasekhar:1985kt}, we work with the PNDs
\begin{align}\label{null-vectors-Kerr}
l^\mu &= \frac{1}{\Delta} \big( r^2 + a^2 , \Delta,\, 0 \, ,\, a \big) \,, \\
\hspace{2cm}
k^\mu & = \frac{1}{2\rho^2} \big( r^2+a^2 , -\Delta, \, 0 \, ,\, a \big) \,, \label{null-vectors-Kerr-k} \\
m^\mu &= \frac{1}{\sqrt{2}(r+i a \cos\theta)} \big( i a \sin\theta , \, 0 \, , \, 1, 
\, i\csc\theta \big) \,\label{null-vectors-Kerr-m}.
\end{align}
The only non-zero component of the Weyl scalars is given by
\begin{equation}\label{Weyl-Kerr}
{\boldsymbol \Psi}_2 = -\frac{M}{(r-i a \cos\theta)^3} \,.
\end{equation}
such that the seed geometry~\eqref{ds2-Kerr} is of Petrov type D, as expected.\footnote{Note that we could also work with other PNDs like \cite{GomezLopez:2017kcw}
	\begin{eqnarray}\label{null-vectors-Kerr-old}
	l^\mu = \frac{1}{\sqrt{2\Delta}\rho} \Big( \rho^2 \sqrt{\Xi}, 
	- \Delta , 0 , \frac{2Mra}{\rho^2\sqrt{\Xi}} \Big) , 
	\hspace{.2cm}
	k^\mu = \frac{1}{\sqrt{2\Delta}\rho} \Big( \rho^2 \sqrt{\Xi}, 
	\Delta ,0 , \frac{2Mra}{\rho^2\sqrt{\Xi}} \Big) , 
	\hspace{.2cm}
	m^\mu = \frac{1}{\sqrt{2}\rho} \Big( 0 , 0 , -1 , \frac{ i \, \csc\theta}{\sqrt{\Xi}} \Big) .
	\end{eqnarray}
	For the above PNDs, all Weyl scalars ${\boldsymbol \Psi}_I$ do not vanish while we saw that only ${\boldsymbol \Psi}_2$ does not vanish when we used the PNDs~\eqref{null-vectors-Kerr}-\eqref{null-vectors-Kerr-m}. Of course, the final results for the Petrov type is independent of the choice of the basis. However, working with the PNDs~\eqref{null-vectors-Kerr} is much easier in practice. Especially for our purpose in this paper, as we will show, working in basis~\eqref{null-vectors-Kerr-old} we have to compute Weyl scalars up to quadratic order for the disformal parameter ${\cal O}(B_0^2)$ while we only need to perform the analysis up to linear order ${\cal O}(B_0)$ in PNDs~\eqref{null-vectors-Kerr}. This extremely simplifies our calculations.}

\subsubsection{Disformed kinematics}

For the disformal vector field, we consider the same configuration as the static spherically symmetric case such that the co-vector field $V_{\mu}\rd x^{\mu}$ has non-vanishing components only along the time and radial direction, namely
\begin{equation}\label{Kerr-V-mu}
V_\mu = \big( V_t, V_r, 0, 0 \big) \,, \hspace{1cm} 
V^\mu = \frac{\rho^2}{\Delta}
\left(-\Xi V_t , \frac{\Delta ^2}{\rho^4}  V_r, 0,-\frac{2 a M r}{\rho^4} V_t\right) 
\,,
\end{equation}
where $V_t:=V_t(t,r)$ and $V_r:=V_r(t,r)$ and we have introduced the notation
\begin{equation}
\Xi \equiv \frac{\Delta}{\rho^2} + \frac{2Mr}{\rho^2} 
\Big( \frac{\Delta}{\rho^2} + \frac{2Mr}{\rho^2}\Big) \,.
\end{equation}
The tetrad components of the vector field can be obtained by substituting Eqs.~\eqref{Kerr-V-mu} and~\eqref{null-vectors-Kerr}-\eqref{null-vectors-Kerr-m} in Eq.~\eqref{Va} as follows 
\begin{equation}\label{Kerr-V-a}
V_a = \left( \frac{r^2+a^2}{\Delta } V_t +V_r , 
\frac{\Delta}{2\rho^2} \Big( \frac{r^2+a^2}{\Delta} V_t - V_r \Big) , 
\frac{i a \sin\theta\, V_t}{\sqrt{2} (r+i a \cos\theta)} , 
-\frac{i a \sin\theta\, V_t}{\sqrt{2} (r-i a \cos\theta)} \right) \,,
\end{equation}
and $V^a$ can be obtained from the above result simply by using $V^a=\eta^{ab}V_b$. 
We see that $V_3=V^2\neq0$, and $V^3=V_2\neq0$ and, such that the disformal co-vector $V_\mu dx^{\mu}$ is not orthogonal to the 2-dimensional surface ${\cal S}$. Notice that this situation corresponds to the disformal transformation used in~\cite{BenAchour:2020fgy} to constructed the disformed Kerr black hole, where the scalar profile depends only on time and radius but not on the $\theta$ angle. See Eq~(3.14) in~\cite{BenAchour:2020fgy}.

Using the above result, the disformed Kerr metric takes the compact form
\begin{eqnarray}\label{ds2-Kerr-dis}
{\widetilde ds}^2_{\rm{Kerr}} = {ds}^2_{\rm{Kerr}} + B_0 \big( V_t dt + V_r dr \big)^2  \,,
\end{eqnarray}
and the disformed PNDs read
\begin{eqnarray}\label{null-vectors-Kerr-dis}
&&{\tilde l}^\mu = l^\mu - \frac{B_0 }{2 }
\left( \frac{r^2+ a^2 }{\Delta} V_t+V_r\right) V^\mu \,, \hspace{1cm} \\
&&{\tilde k}^\mu = k^\mu - \frac{B_0}{4} 
\left( 
\frac{r^2+ a^2 }{\Delta} V_t-V_r\right) \frac{\Delta}{\rho^2} V^\mu
\,,\label{null-vectors-Kerr-dis-k}\\
&&{\tilde m}^\mu = m^\mu - \frac{B_0 }{2\sqrt{2}}
\frac{i a \sin\theta\, V_t }{r+i a \cos\theta} V^\mu
\,,\label{null-vectors-Kerr-dis-m}
\end{eqnarray}
where the explicit forms of undisformed PNDs and disformal vector field are given by Eqs.~\eqref{null-vectors-Kerr}-\eqref{null-vectors-Kerr-m} and~\eqref{Kerr-V-mu}. Note that, contrary to the case of static spherically symmetric case presented in~\eqref{null-vectors-SS-dis} where i) $V\in {T}_p^\star({\cal M}/{\cal S})$ and ii) the PNDs ${\tilde m}^\mu$ and $\tilde{\bar m}^\mu$ did not change, in this axisymmetric case, $V\notin {T}_p^\star({\cal M}/{\cal S})$ and all PNDs change after performing disformal transformation.

In the previous example of static spherically symmetric solution, we have computed all nonzero scalars in Eqs.~\eqref{Dis-Scalars-SS} and~\eqref{Psi-SS}. Here, we only compute those scalars that we need, illustrating if needed the efficiency of our method. We first notice that working in the null basis~\eqref{null-vectors-Kerr}-\eqref{null-vectors-Kerr-m}, all ${\boldsymbol \Psi}_I$ vanish except ${\boldsymbol \Psi}_2$.
Using this result, one can show that the first non-zero contribution to ${\tilde D}$ shows up at the second order ${\cal O}(B_0^2)$ as 
\begin{equation}\label{Dtilde}
\tilde{ D} = 81 {\boldsymbol \Psi}_2^2 {\boldsymbol \Delta}_0 {\boldsymbol \Delta}_4 \,,
\end{equation}
which demonstrates that ${\tilde D}\neq0$ for ${\boldsymbol \Delta}_0\neq0$ and ${\boldsymbol \Delta}_4\neq0$. 

Before concluding, we point that since ${\tilde D}={\cal O}(B_0^2)$ for ${\boldsymbol \Delta}_I={\cal O}(B_0)$, a legitimate question would be whether one can trust this analysis up to linear order ${\cal O}(B_0)$ for $\tilde{\boldsymbol \Psi}_I$? Interestingly, it turns out that one can easily show that the result~\eqref{Dtilde} holds in the null basis~\eqref{null-vectors-Kerr}-\eqref{null-vectors-Kerr-m} even if one computes all $\tilde{\boldsymbol \Psi}_I$ up to the quadratic order ${\cal O}(B_0^2)$. Had we worked with another null basis, for instance the null basis~\eqref{null-vectors-Kerr-old}, we would had to keep calculations for $\tilde{\boldsymbol \Psi}_I$ up to the second order ${\cal O}(B_0^2)$ which would make the calculations much more involved. Thus, working with the basis~\eqref{null-vectors-Kerr}-\eqref{null-vectors-Kerr-m} allows us to only compute ${\boldsymbol \Delta}_0$  and ${\boldsymbol \Delta}_4$ up to the linear order ${\cal O}(B_0)$. The explicit expressions are complicated and we only present the results for $\theta=\pi/2$ which are given by
\begin{eqnarray}\label{kerr-Delta0}
{\boldsymbol \Delta}_0 &=& \frac{B_0 a^2}{2 r^4 } \Bigg\{
V_r^2 + \frac{4 M^2 r^2}{\Delta^2} V_t^2
+ r^2 \left(\dot{V}_r^2+ V_t'{}^2-\dot{V}_r V_t'-\dot{V}_t V_r'\right) \nonumber
\\ \nonumber
&+& \left( r \big( \ddot{V}_r - \dot{V}_t' \big)
+2 \frac{a^2+r^2}{\Delta} \dot{V}_t
+ 2 \dot{V}_r + \frac{4 M}{\Delta} V_t \right) r V_r 
+ 2 \frac{a^2+r^2}{\Delta} r \dot{V}_r V_t
\\
&+&\Big( r \big(V_t''-\dot{V}_r\big)-2 V_t' \Big) r V_t
- \frac{a^2+3 r^2}{\Delta} \bigg( \dot{V}_t
- \frac{a^2+r^2}{\Delta} \dot{V}_t \bigg) r V_t \Bigg\}
\,, \\ \label{kerr-Delta4}
{\boldsymbol \Delta}_4 &=& \frac{\Delta^2}{4r^4} \left\{ {\boldsymbol \Delta}_0 
-\frac{2 B_0 a^2}{r^4}
\left(\frac{M r}{\Delta} \frac{a^2+3 r^2}{\Delta} V_t \dot{V}_t
+ \left(\frac{2 M}{\Delta} V_t+  r \dot{V}_r\right) V_r \right) \right\} \,.
\end{eqnarray}
From Table \ref{tab1}, the above results show that the Petrov type of the stealth Kerr solution~\eqref{ds2-Kerr} changes from type D to type I. This example explicitly confirms that not only PNDs but also Petrov type of a spacetime can change after performing a disformal transformation. As a result, the non-stealth Kerr-like black hole solution derived in~\cite{BenAchour:2020fgy, Anson:2020trg} for DHOST theories is found to be algebraically general of type I. We have summarized the results of this section in Table \ref{tab2}.

The fact that the Petrov type of black hole solutions changes in the context of modified gravity theories was already noticed in different contexts, such as dynamical Chern-Simons and Gauss-Bonnet gravity~\cite{Yagi:2012ya,Owen:2021eez}. While the result we have obtained for the stealth Kerr parallels these results (where the slowly rotating black holes solutions analyzed there go from type D to type I too), our analysis turns out to be conceptually more subtle. Indeed, because disformal transformation reduces to field redefinition in the absence of matter fields coupled to the gravitational sector, it might seem awkward to find different Petrov type for the stealth Kerr and its disformed version. However, gravity is generated and is probed by the matter fields. Therefore, the key point is that determining the Petrov type of a given solution requires the introduction of a frame in which matter fields (including the standard model of particle physics) are minimally coupled to the metric, if one would like the classification based on the Petrov type to be physically relevant. In this frame free-falling observers follow geodesics and electromagnetic waves propagate along lightcones, while in other frames they do not in general. In this situation the Petrov classification can characterize useful physical properties of the solution measured by such observers and electromagnetic waves if and only if the classification is associated with the Weyl scalars computed in the frame where matter fields are minimally coupled to the metric. Thus, when we deal with the disformal transformation as a field redefinition which includes metric, we have to determine to which frame the matter fields are minimally coupled. Depending on this choice, gravity shows different properties. As we emphasized in the first section, the Petrov classification is performed by introducing a null tetrad which corresponds to the frame of an implicit observer moving on the underlying geometry. This step can be understood as introducing a disguised test matter field and specifying its coupling to the metric. For this reason,  as a gravitational properties of the system, the Petrov type is not invariant under disformal transformation and the analysis of the stealth Kerr metric and its disformed version illustrates this fact.
\begin{table}
	\centering
	\begin{tabular}{ | p{6cm}|p{3cm}|p{3cm}  | }
		\hline
		\multicolumn{3}{|c|}{\bf Petrov type after disformal transformation (DT)} \\
		\hline 
		\hfil Solution & \hfil Before DT & \hfil After DT \\
		\hline
		\hfil FLRW Eq.~\eqref{ds2-FLRW} & \hfil O & \hfil O \\
		\hfil Spherically Symmetric Eq.~\eqref{ds2-SS} & \hfil D & \hfil D \\
		\hfil Kerr Eq.~\eqref{ds2-Kerr} & \hfil D & \hfil I \\
		\hline
	\end{tabular}
	\newline
	\caption{Petrov types of the FLRW background and static spherically symmetric solutions do not change after performing a disformal transformation. However, Petrov type of the Kerr black hole changes from type D to type I after performing a disformal transformation and it is no longer algebraically special.}\label{tab2}
\end{table}

\section{Summary and discussion}\label{sec5}

We have presented a detailed analysis allowing to capture how the Petrov type of a given seed solution is modified under a disformal transformation. As a first step, we have introduced the map (\ref{T-map}) relating the null tetrad in the Einstein frame to its counterpart in the Jordan frame. Using this map, we have then derived the close formulas (\ref{kl-disform}) relating the null tetrads before and after disformal transformations, providing key tools to understand how the causal structure of a given seed changes under such mapping. As a second step, we have decomposed the Weyl tensor in a suitable form given by (\ref{Weyl-JE-Sym}), paying special attention to the symmetries of the different contributions. From that result, we have then obtained explicit formulas relating the Weyl scalars before and after the disformal transformations, which provide the key relations to understand the fate of the Petrov type under disformal field redefinitions. See (\ref{Psi0})-(\ref{Psi4}).

Subsequently, we have applied these general formulas to a specific setup, namely the case of simplified pure disformal transformations, allowing the derivation of simple enough expressions for the disformed Weyl scalars. Finally, we have applied our method to three relevant examples of seed solutions discussed in the literature:

\begin{itemize}
	\item We have first considered an FLRW cosmological background~\eqref{ds2-FLRW} which is of Petrov type O. We have shown that performing a disformal transformation with a homogeneous and timelike vev~\eqref{V-FLRW} for the disformal vector, the Petrov type of the disformed metric~\eqref{ds2-FLRW-dis} does not change. The result is consistent with the well-known fact that all FLRW spacetimes are conformally flat and thus of Petrov type O. 
	
	\item As a second example, we have focused on a general static spherically symmetric backgrounds~\eqref{ds2-SS} which are Petrov type D. We have found that for such backgrounds and for the specific vector profile~\eqref{SS-V-mu}, the PNDs~\eqref{null-vectors-SS-dis} and~\eqref{null-vectors-SS-dis-k} change while the Petrov type remains type D. The result is consistent with the well-known fact that all spherically symmetric spactimes are of Petrov type D (or O). 
	
	\item Finally, we have considered the seed solution consisting in the stealth rotating Kerr background~\eqref{ds2-Kerr}. Performing a disformal map with vector profile given by~\eqref{Kerr-V-mu}, which includes the case investigated in~\cite{BenAchour:2020fgy, Anson:2020trg} to construct the disformed Kerr black hole, we have found that not only the PNDs change~\eqref{null-vectors-Kerr-dis}-\eqref{null-vectors-Kerr-dis-m} but the Petrov type also changes from the algebraically special type D to algebraically general type I. 
\end{itemize}

From a technical point of view, the formulas provided in this work shall allow one to identify in a straightforward manner if and how the Petrov type of a seed solution will be modified when performing a disformal field redefinition to construct new exact solution in scalar-tensor and vector-tensor theories. Moreover starting from the general conditions of invariance of the Petrov type, one can also design a suitable disformal transformation in order to construct a given new algebraically special solution. In that regard, this possibility extends the standard use of the Petrov classification to construct exact solutions.

From a more general perspective, the knowledge of the Petrov type of a given solution reveals key information on its geometric structure. The fact that the non-stealth Kerr solution is of Petrov type I shows that the nature of its PNDs radically differs from its GR counterpart. While this solution is stationary and does not carry propagating gravitational waves, one can nevertheless characterize it by investigating its multipole expansion. It would provide an interesting way to compare the two geometries. While the characterization of the multipolar moments of a given geometry is not free from ambiguities~\cite{Bonga:2021ouq}, it can nevertheless be worked out in a clean manner for stationary, axisymmetric and asymptotically flat geometries following the Geroch-Hansen approach~\cite{Geroch, Hansen, Beig}. Whether this approach can be used to further capture the structural differences between the Kerr geometry and its disformed version constructed in~\cite{BenAchour:2020fgy, Anson:2020trg} provides an interesting direction which we plan to investigate in the near future.
\vspace{0.7cm}

{\bf Acknowledgments:} The work of J.~BA is supported by the Alexander von Humboldt foundation.  The work of A.D.F.\ was supported by Japan Society for the Promotion of Science Grants-in-Aid for Scientific Research
No.\ 20K03969. The work of M.A.G.\ was supported by Japan Society for the Promotion of Science (JSPS) Grants-in-Aid for international research fellow No.\ 19F19313. 
The work of S.M was supported in part by Japan Society for the Promotion of Science Grants-in-Aid for Scientific Research No.~17H02890, No.~17H06359, and by World Premier International Research Center Initiative, MEXT, Japan. M.C.P.\ acknowledges the support from the Japanese Government (MEXT) scholarship for Research Student.


\vspace{0.5cm}

\appendix

\section{Tetrad representation}\label{app-tetrad-rep}
\setcounter{equation}{0}
\renewcommand{\theequation}{A\arabic{equation}}

\subsection{Riemann tensor in Einstein frame}
In this appendix we define appropriate real and complex scalars which present all tetrad components of the Riemann tensor. The Riemann tensor $R_{\alpha\beta\mu\nu}$ can be decomposed into the trace and trace-free parts according to the well-known Weyl decomposition
\begin{equation}\label{Riemann}
{R}_{\alpha\beta\mu\nu} =
{C}_{\alpha\beta\mu\nu} - 
\frac{1}{2} \big( {R}_{\alpha\nu} { g}_{\beta\mu} + { R}_{\beta\mu} g_{\alpha\nu} 
- {R}_{\alpha\mu} { g}_{\beta\nu} - { R}_{\beta\nu} g_{\alpha\mu} \big) 
- \frac{1}{6} { R} \big( g_{\alpha\mu} { g}_{\beta\nu} - g_{\alpha\nu} { g}_{\beta\mu} \big) \,,
\end{equation}
where $C_{\alpha\beta\mu\nu}$ is the Weyl part which is trace-free with respect to any index, $R_{\mu\nu} = g^{\alpha\beta}R_{\alpha\mu\beta\nu}$ is the Ricci tensor, and $R = g^{\alpha\beta} R_{\alpha\beta}$ is the Ricci scalar. We note that the above geometric decomposition is valid for any tensor with the same symmetries as the Riemann tensor. In tetrad representation, the tetrad components of the Riemann tensor are given by
\begin{equation}\label{Riemann-T}
{R}_{abcd} =
{C}_{abcd} - 
\frac{1}{2} \big( {R}_{ad} { \eta}_{bc} + { R}_{bc} \eta_{ad} 
- {R}_{ac} { \eta}_{bd} - { R}_{bd} \eta_{ac} 
\big) 
- \frac{1}{6} { R} \big( \eta_{ac} { \eta}_{bd} - \eta_{ad} { \eta}_{bc} \big) \,,
\end{equation}
where $R_{abcd} = R_{\alpha\beta\mu\nu} \theta^\alpha{}_a \theta^\beta{}_b \theta^\mu{}_c \theta^\nu{}_d$ and so on. $\theta^{\mu}{}_a$ are any tetrad basis. The Riemann tensor $R_{abcd}$ has twenty tetrad components and, in the above decomposition, ten components are encoded in the Weyl part $C_{abcd}$ and the remaining ten components are encoded in the Ricci part $R_{ab}$. 

In the null tetrad basis defined by Eq.~\eqref{T}, we can express ten components of the Weyl part by means of the five complex Weyl scalars ${\boldsymbol \Psi}_I$ as follows
\begin{eqnarray}\label{WeylS-T-a}
{\boldsymbol \Psi}_0 = C_{0202} \,, \hspace{.5cm}
{\boldsymbol \Psi}_1 = C_{0102}  \,, \hspace{.5cm}
{\boldsymbol \Psi}_2 = C_{0231} \,, \hspace{.5cm}
{\boldsymbol \Psi}_3 = C_{1013} \,, \hspace{.5cm}
{\boldsymbol \Psi}_4 = C_{1313} \,.
\end{eqnarray}
As it is clear from relations~\eqref{WeylS-T-a}, the Weyl scalars $\Psi_I$ are some special tetrad components of the Weyl tensor $C_{abcd}$ which describe ten independent components of the Weyl tensor. The other tetrad components of the Weyl tensor are not independent of the Weyl scalars and their complex conjugates and we have~\cite{Chandrasekhar:1985kt}\footnote{Note that all the scalar quantities that we define in this paper are different than those defined in~\cite{Chandrasekhar:1985kt} only by a minus sign and we use the different metric signature with the most positive signs.}
\begin{eqnarray}\label{Weyl-components}
&& C_{0203} = C_{1213} = C_{0221} = C_{0331} = 0 \,,
\\ \nonumber 
&&C_{0120} = C_{0223}\,, \hspace{1cm} 
C_{0130} = C_{0332}\,, \hspace{1cm} 
C_{0121} = C_{1232}\,, \hspace{1cm} 
C_{0131} = C_{1323}\,, 
\\ \nonumber 
&&C_{0101} = C_{2323}\,, \hspace{1cm} 
C_{0231} = \frac{1}{2} ( C_{0101} - C_{0123} ) = \frac{1}{2} ( C_{2323} - C_{0123} ) \,,
\\ \nonumber 
&& C_{0101} = {\boldsymbol \Psi}^\ast_2 + {\boldsymbol \Psi}_2 \,, \hspace{1cm} 
C_{0123} = {\boldsymbol \Psi}^\ast_2 - {\boldsymbol \Psi}_2 \,.
\end{eqnarray}
Other components can be easily obtained by using the symmetries of the Weyl tensor and also taking complex conjugate. For example, $C_{0312} = C^*_{0213} = - C^*_{0231}=-{\boldsymbol \Psi}^*_2$. Note that taking complex conjugate indices $0$ and $1$ does not change while indices $2$ and $3$ exchange with each other. The reason is that indices $0$ and $1$ correspond to the real null tetrad components $l^\mu$ and $k^\mu$ while indices $2$ and $3$ correspond to the complex null tetrad components $m^\mu$ and ${\bar m}^\mu$.

Also ten components of the Ricci part can be expressed in terms of four real scalars ${\boldsymbol \Phi}_{00}$, ${\boldsymbol \Phi}_{11}$, ${\boldsymbol \Phi}_{22}$, ${\boldsymbol \Lambda}$, and three complex scalars ${\boldsymbol \Phi}_{01}$, ${\boldsymbol \Phi}_{02}$, ${\boldsymbol \Phi}_{12}$ which are defined as
\begin{eqnarray}\label{RicciS-T-a}
&&{\boldsymbol \Phi}_{00} = \frac{1}{2} R_{00} \,, \hspace{.5cm}
{\boldsymbol \Phi}_{11} = \frac{1}{4} (R_{01}+R_{23}) \,, \hspace{.5cm}
{\boldsymbol \Phi}_{22} = \frac{1}{2} R_{11} \,, \hspace{.5cm}
{\boldsymbol \Lambda} = - \frac{1}{24} R = \frac{1}{12} (R_{01} - R_{23}) \,, \nonumber \\
&&{\boldsymbol \Phi}_{01} = \frac{1}{2} R_{02} \,, \hspace{1.5cm}
{\boldsymbol \Phi}_{02} = \frac{1}{2} R_{22} \,, \hspace{1.5cm}
{\boldsymbol \Phi}_{12} = \frac{1}{2} R_{12} \,.
\end{eqnarray}
We can express all components of the Ricci tensor in terms of the above scalar quantities and their complex conjugates. For example, $R_{01} = 2 ({\boldsymbol \Phi}_{11}+3{\boldsymbol \Lambda})$, $R_{23} = 2 ({\boldsymbol \Phi}_{11}-3{\boldsymbol \Lambda})$, $R_{03} = 2 {\boldsymbol \Phi}^\ast_{01}$, $R_{13}= 2 {\boldsymbol \Phi}^\ast_{12}$, and $R_{33} = 2 {\boldsymbol \Phi}^\ast_{02}$.
Here we considered the Riemann tensor in the Einstein frame while the results is true for the Riemann tensor in the Jordan frame ${\tilde R}_{abcd}$ and also any other tensor with the same symmetry group as the Riemann tensor.

\subsection{Weyl tensor in Jordan frame}

Let us use the results of the previous subsection and express the Weyl tensor in Jordan frame ${\tilde C}_{abcd}$ given by Eq.~\eqref{C-tilde-trans0} in terms of the Einstein frame quantities after disformal transformation. For the l.h.s.\ of Eq.~\eqref{C-tilde-trans0}, as we mentioned above, we can simply express all components of ${\tilde C}_{abcd}$ in terms of the five complex Weyl scalars in the Jordan frame similar to Eq.~\eqref{WeylS-T-a} as follows
\begin{eqnarray}\label{WeylS-tilde-T-a}
\tilde{\boldsymbol \Psi}_0 = { \tilde C}_{0202} \,, \hspace{.5cm}
\tilde{\boldsymbol \Psi}_1 = { \tilde C}_{0102}  \,, \hspace{.5cm}
\tilde{\boldsymbol \Psi}_2 = { \tilde C}_{0231} \,, \hspace{.5cm}
\tilde{\boldsymbol \Psi}_3 = { \tilde C}_{1013} \,, \hspace{.5cm}
\tilde{\boldsymbol \Psi}_4 = { \tilde C}_{1313} \,,
\end{eqnarray}
and the other components can be found with relations similar to~\eqref{Weyl-components}.

For the r.h.s.\ of Eq.~\eqref{C-tilde-trans0}, ${ C}_{abcd}$ can be expressed in terms of the five complex Weyl scalars~\eqref{WeylS-T-a}. Since the tensor $B_{\alpha\beta\mu\nu}$ has the same symmetry as the Weyl tensor, $B_{abcd}$ can be expressed in terms of five complex Weyl-type scalars as follows
\begin{eqnarray}\label{Delta-a}
{\boldsymbol \Delta}_0 = B_{0202} \,, \hspace{.5cm}
{\boldsymbol \Delta}_1 = B_{0102} \,, \hspace{.5cm}
{\boldsymbol \Delta}_2 = B_{0231} \,, \hspace{.5cm}
{\boldsymbol \Delta}_3 = B_{1013} \,, \hspace{.5cm}
{\boldsymbol \Delta}_4 = B_{1313} \,,
\end{eqnarray}
where the other components can be obtained through the same relations as~\eqref{Weyl-components} for $B_{abcd}$. The remaining part is $Z^S_{abcd}$ which from Eq.~\eqref{Z-S} we find
\begin{eqnarray}\label{Zij-a}
Z^S_{abcd} =  
-\frac{1}{2} \big(Z_{ad} \eta_{bc} + Z_{bc} \eta_{ad}
- Z_{ac} \eta_{bd} - Z_{bd} \eta_{ac} \big) 
- \frac{1}{6} Z (\eta_{ac} \eta_{bd} - \eta_{ad} \eta_{bc}) \,.
\end{eqnarray}
Hence $Z^S_{abcd}$ is completely expressed in terms of the Ricci part $Z_{ab}$ and, therefore, it can be completely expressed in terms of the four real and three complex scalars similar to relations~\eqref{RicciS-T-a}. We then define 
\begin{eqnarray}\label{Pi-a}
&&{\boldsymbol \Pi}_{00} = \frac{1}{2} Z_{00} \,, \hspace{.5cm}
{\boldsymbol \Pi}_{11} = \frac{1}{4} (Z_{01}+Z_{23}) \,, \hspace{.5cm}
{\boldsymbol \Pi}_{22} = \frac{1}{2} Z_{11} \,, \hspace{.5cm}
{\boldsymbol \Lambda}^S = - \frac{1}{24} Z = \frac{1}{12} (Z_{01} - Z_{23}) \,, \nonumber \\
&&{\boldsymbol \Pi}_{01} = \frac{1}{2} Z_{02} \,, \hspace{1.5cm}
{\boldsymbol \Pi}_{02} = \frac{1}{2} Z_{22} \,, \hspace{1.5cm}
{\boldsymbol \Pi}_{12} = \frac{1}{2} Z_{12} \,.
\end{eqnarray}
We can express all tetrad components of $Z^S_{abcd}$ in terms of the above defined scalars. Some of them are  as follows
\begin{eqnarray}\label{Z-P}
&&Z^S_{0101} = 2 ({\boldsymbol \Lambda}^S+{\boldsymbol \Pi}_{11}) \,, \hspace{.5cm}
Z^S_{0102} = {\boldsymbol \Pi}_{01} \,, \hspace{.5cm}
Z^S_{0112} = - {\boldsymbol \Pi}_{12} \,, \hspace{.5cm}
Z^S_{0123} = 0 \,,
\nonumber \\ 
&&Z^S_{0203} = {\boldsymbol \Pi}_{00} \,, \hspace{.5cm}
Z^S_{0212} = - {\boldsymbol \Pi}_{02} \,, \hspace{.5cm}
Z^S_{0213} = 2 {\boldsymbol \Lambda}^S \,, \hspace{.5cm}
Z^S_{0223} = {\boldsymbol \Pi}_{01} \,,
\nonumber \\
&&Z^S_{1212} = 0 \,, \hspace{.5cm}
Z^S_{1213} = {\boldsymbol \Pi}_{22} \,, \hspace{.5cm}
Z^S_{1223} = {\boldsymbol \Pi}_{12} \,, \hspace{.5cm}
Z^S_{2323} =  2 ({\boldsymbol \Lambda}^S-{\boldsymbol \Pi}_{11}) \,.
\end{eqnarray}
Now, from Eq.~\eqref{C-tilde-trans} we can express $\tilde{\boldsymbol \Psi}_I$ in terms of ${\boldsymbol \Psi}_I$, ${\boldsymbol \Delta}_I$, ${\boldsymbol \Pi}_{IJ}$, and ${\boldsymbol \Lambda}^S$ as follows
\begin{eqnarray}\label{Psi0}
A \tilde{\boldsymbol \Psi}_0 &=&  \gamma^2 ( {\boldsymbol \Psi}_0 + {\boldsymbol \Delta}_0 )
+ \beta^2 \big[ ( {\boldsymbol \Psi}^*_0 + {\boldsymbol \Delta}^*_0 ) (V_2V_2)^2 
+ ( {\boldsymbol \Psi}^*_4 + {\boldsymbol \Delta}^*_4 ) (V^1V^1)^2 \big]
\nonumber \\ 
&+& 2 \beta^2 V^1 V_2 \big[ 2 ( {\boldsymbol \Psi}^*_1 + {\boldsymbol \Delta}^*_1 ) V_2V_2 
+ 3 ( {\boldsymbol \Psi}^*_2 + {\boldsymbol \Delta}^*_2 ) V^1V_2 
+ 2 ( {\boldsymbol \Psi}^*_3 + {\boldsymbol \Delta}^*_3 ) V^1V^1 \big] 
\nonumber \\ 
&-& 2 \frac{B}{A} \beta \gamma \big[ 2 {\boldsymbol \Pi}_{01} V^1V_2 
+ {\boldsymbol \Pi}_{02} V^1 V^1 + {\boldsymbol \Pi}_{00} V_2 V_2 \big]
\,,
\end{eqnarray}

\begin{eqnarray}\label{Psi1}
A \tilde{\boldsymbol \Psi}_1 &= &
\gamma \big[ ({\boldsymbol \Psi} _1+{\boldsymbol \Delta} _1) \left( 1 - 2 \beta V^1 V_1 \right)
- \beta  V^2 V_1 ({\boldsymbol \Psi} _0+{\boldsymbol \Delta} _0)
+ \beta  V^1 V_2 ({\boldsymbol \Psi} _2+{\boldsymbol \Delta} _2) \big]
\nonumber \\ 
&-& \beta  \big(
V_2 V_2 \left( {\boldsymbol \Psi}^*_1+ {\boldsymbol \Delta}^*_1\right)
+ 2 V^1 V_2 \left({\boldsymbol \Psi}^*_2 + {\boldsymbol \Delta}^*_2\right)
+ V^1 V^1 \left({\boldsymbol \Psi}^*_3+{\boldsymbol \Delta}^*_3\right)
\big)
\nonumber \\ 
&+& \beta^2 \big[ 
V_1 V_2 V_2
\big(
V_2 ( {\boldsymbol \Psi}^*_0 + {\boldsymbol \Delta}^*_0 ) 
+ 4 V^1 \left({\boldsymbol \Psi}^*_1+{\boldsymbol \Delta}^*_1\right)
\big)
- V^2 V^1V^1V^1 \left({\boldsymbol \Psi}^*_4+{\boldsymbol \Delta}^*_4\right) 
\nonumber \\ 
&&
+ 2 V^1V^1 \left({\boldsymbol \Psi}^*_3+{\boldsymbol \Delta}^*_3 \right) \left(V^1 V_1-V^2 V_2\right)
+V^1 V_2 \left({\boldsymbol \Psi}^*_2+{\boldsymbol \Delta}^*_2\right) \left(5 V^1 V_1- V^2 V_2\right) 
\big]
\nonumber \\ 
&+& \frac{B}{A} \Big\{
{\boldsymbol \Pi}_{01} \big[ 
\gamma \left( 1 - 2 \beta V^1 V_1 \right) + 2 \beta^2 V_1V^1 V_2 V^2
\big]
\nonumber \\ 
&-& \beta
\big[ 2 V^1 V_2 \left({\boldsymbol \Pi} _{11}+ 2 {\boldsymbol \Lambda}^S\right)
+\Pi_{12} V^1 V^1 - {\boldsymbol \Pi}_{02} V^2 V^1 
+ V_2 \left( {\boldsymbol \Pi}^*_{01} V_2 + {\boldsymbol \Pi} _{00} V_1\right)\big]
 \\ \nonumber 
&+& \beta^2 \big[ 
2 V_1 V_2 \big(2 V^1 V^1 \left({\boldsymbol \Pi}_{11}+{\boldsymbol \Lambda}^S\right)
+{\boldsymbol \Pi}^*_{01} V^1 V_2 + {\boldsymbol \Pi}_{00} V^2V_2 \big)
+ 2 {\boldsymbol \Pi} _{12} V_1 V^1V^1V^1
 \\ \nonumber 
&&
-  V^1 V_2 \left( {\boldsymbol \Pi}_{00} V_1 V^1 + {\boldsymbol \Pi}_{02} V^2 V^2 
+ {\boldsymbol \Pi}_{22} V^1 V^1 
+2 {\boldsymbol \Pi}^*_{12} V^1V_2
+{\boldsymbol \Pi}^*_{02} V_2 V_2\right) - 4 V^2 V_2 {\boldsymbol \Lambda}^S
\big]
\Big\} \,,
\end{eqnarray}

\begin{eqnarray}
\label{Psi2}
A \tilde{\boldsymbol \Psi}_2 &=& \gamma^2 ( {\boldsymbol \Psi}_2 + {\boldsymbol \Delta}_2 )
+ \beta^2 \big[ ( {\boldsymbol \Psi}^*_0 + {\boldsymbol \Delta}^*_0) (V_1 V_2)^2 
+ ( {\boldsymbol \Psi}^*_4 + {\boldsymbol \Delta}^*_4 ) (V^1 V^2)^2 \big]
\nonumber \\ 
&+& \beta^2 \big[ 
2 (V_1V^1-V_2V^2) \big( ( {\boldsymbol \Psi}^*_1 + {\boldsymbol \Delta}^*_1 ) V_1 V_2 
- ( {\boldsymbol \Psi}^*_3 + {\boldsymbol \Delta}^*_3 ) V^1 V^2 \big)
\nonumber \\ 
&&+ ( {\boldsymbol \Psi}^*_2 + {\boldsymbol \Delta}^*_2 ) 
\big( (V_1V^1-V_2 V^2)^2 - 2 V_1 V^1 V_2 V^2 \big)
\big] 
- 2 \frac{B}{A} {\boldsymbol \Lambda}^S \big[ 1 + 2 \beta \gamma (1-\gamma) \big]
\\ \nonumber
&-&
\frac{B}{A} \beta \gamma \big[ {\boldsymbol \Pi}_{00} V_1 V_1 - 2 {\boldsymbol \Pi}_{01} V_1 V^2 
+ {\boldsymbol \Pi}_{02} V^2 V^2 
+ 2 {\boldsymbol \Pi}^*_{12} V^1 V_2 
+ {\boldsymbol \Pi}^*_{02} V_2 V_2 + {\boldsymbol \Pi}_{22} V^1 V^1 \big] \,,
\end{eqnarray}

\begin{eqnarray}\label{Psi3}
A \tilde{\boldsymbol \Psi}_3 & = & \gamma \big[
({\boldsymbol \Psi}_3 + {\boldsymbol \Delta}_3) \left(1-2 \beta V^1 V_1\right) 
- \beta ({\boldsymbol \Psi}_2+ {\boldsymbol \Delta}_2) V^2 V_1 
+ \beta ({\boldsymbol \Psi}_4+{\boldsymbol \Delta}_4) V^1 V_2 \big]
\nonumber \\ 
&-& \beta \big[ 
V_1 V_1 \left( {\boldsymbol \Psi}^*_1 + {\boldsymbol \Delta}^*_1\right)
- 2 V^2 V_1 \left( {\boldsymbol \Psi}^*_2 + {\boldsymbol \Delta}^*_2\right)
+ V^2 V^2 \left( {\boldsymbol \Psi}^*_3 + {\boldsymbol \Delta}^*_3\right)
\big]
\nonumber \\ 
& + & \beta^2 \big[ 
2 ( {\boldsymbol \Psi}^*_1 + {\boldsymbol \Delta}^*_1) V_1^2 \left(V^1 V_1 - V^2 V_2\right)
+ V_1 V_1 V_1 V_2 \left( {\boldsymbol \Psi}^*_0+{\boldsymbol \Delta}^*_0\right)
- V^1 V^2 V^2V^2 \left( {\boldsymbol \Psi}^*_4+{\boldsymbol \Delta}^*_4\right)
\nonumber \\ 
&&
+ 4 V^1 V^2 V^2 V_1 \left( {\boldsymbol \Psi}^*_3+{\boldsymbol \Delta}^*_3\right)
+ ( {\boldsymbol \Psi}^*_2 + {\boldsymbol \Delta}^*_2) V^2 V_1 \left(V^2 V_2-5 V^1 V_1\right)
\big]
\nonumber \\ 
&+& \frac{B}{A} \Big\{ 
{\boldsymbol \Pi}^*{}_{12} \big[ 
\gamma \left( 1 - 2 \beta V^1 V_1 \right) + 2 \beta^2 V_1V^1 V_2 V^2
\big]
\nonumber \\ 
&-& \beta \big[
\left( {\boldsymbol \Pi}_{12} V^2 - {\boldsymbol \Pi}_{22} V^1 \right) V^2
-2 V^2 V_1 \left( {\boldsymbol \Pi} _{11}+2 {\boldsymbol \Lambda}^S\right)
+ {\boldsymbol \Pi}^*_{01} V_1 V_1 + {\boldsymbol \Pi}^*_{02} V_2 V_1
\big]
\nonumber \\ 
&+& \beta^2 \big[
V^2 V_1 \left( {\boldsymbol \Pi}_{02} V^2V^2 + 2 {\boldsymbol \Pi} _{12} V^1 V^2
- {\boldsymbol \Pi}_{22} V^1V^1 + {\boldsymbol \Pi}^*_{02} V_2 V_2 
- 4 V^2 V_2 {\boldsymbol \Lambda}^S \right)
- 2 {\boldsymbol \Pi}_{22} V^1 V^2V^2 V_2
\nonumber \\ 
&&+ V_1 V_1 \big( \left(2 {\boldsymbol \Pi}^*_{01} V^1 + {\boldsymbol \Pi}_{00} V^2\right)  V_1 
- 2
\left(2 V^1 \left( {\boldsymbol \Pi} _{11} + {\boldsymbol \Lambda}^S\right) 
+ {\boldsymbol \Pi}_{01} V^2 \right) V^2 \big)
\big]
\Big\} \,,
\end{eqnarray}

\begin{eqnarray}\label{Psi4}
A \tilde{\boldsymbol \Psi}_4 &=& \gamma^2 ( {\boldsymbol \Psi}_4 + {\boldsymbol \Delta}_4 ) 
+ \beta^2 \big[ ( {\boldsymbol \Psi}^*_4 + {\boldsymbol \Delta}^*_4 ) (V^2V^2)^2 
+ ( {\boldsymbol \Psi}^*_0 + {\boldsymbol \Delta}^*_0 ) (V_1V_1)^2 \big]
\nonumber \\ 
&-& 2 \beta^2 V_1 V^2 \big[ 
2 ( {\boldsymbol \Psi}^*_1 + {\boldsymbol \Delta}^*_1 ) V_1 V_1
-3 ( {\boldsymbol \Psi}^*_2 + {\boldsymbol \Delta}^*_2 ) V_1 V^2
+ 2 ( {\boldsymbol \Psi}^*_3 + {\boldsymbol \Delta}^*_3 ) V^2 V^2 \big] 
\nonumber \\ 
&+& 2 \frac{B}{A} \beta \gamma
\big[  2 {\boldsymbol \Pi}^*_{12} V_1 V^2 - {\boldsymbol \Pi}^*_{02} V_1 V_1 
- {\boldsymbol \Pi}_{22} V^2 V^2 \big]
\,,
\end{eqnarray}
where we have defined parameter
\begin{equation}\label{gamma}
\gamma \equiv 1 - \beta ( V_1V^1+V_2V^2 ) \,.
\end{equation}
The above parameter goes to unity for $B=0$ as $\beta=0$ in this case which can be seen from Eq.~\eqref{beta}.

\end{document}